\documentclass[preprint2]{aastex6}

\usepackage{mathpazo}

\newcommand{\bls}{1.0}
\newcommand{\blsA}{1.0}
\newcommand{\blsC}{1.0}

\newcommand{\myurl}{\url{www2.hawaii.edu/~zeebe/Astro.html}}

\let\bm\relax

\newcommand{\lsim}{\mbox{\raisebox{-.0ex}{$\stackrel{<}{_\sim} \ $}}}
\newcommand{\gsim}{\raisebox{-.5ex}{$\stackrel{>}{\sim} \ $}}
\newcommand{\fns}{\footnotesize}
\newcommand{\rd}{\mbox{\rm d}}

\newcommand{\sm}{\mbox{$\sim$}}
\newcommand{\Dt}{\mbox{$\D t$}}
\newcommand{\cS}{\mbox{${\cal S}$}}
\newcommand{\eE}{\mbox{$e_{\cal E}$}}

\newcommand{\DeE}{\mbox{$\D e_{\cal E}$}}

\newcommand{\mE}{\mbox{$m_{\cal E}$}}

\newcommand{\lE}{\mbox{$\varpi_{\cal E}$}}
\newcommand{\lEp}{\mbox{$\varpi'_{\cal E}$}}
\newcommand{\hnb}{{\tt HNBody}}
\newcommand{\ems}{{\tt ems}}
\newcommand{\lun}{{\tt lun}}
\newcommand{\as}{\arcsec}
\newcommand{\asy}{{\as}yr\pmo}

\newcommand{\D}{\Delta}
\newcommand{\e}[1]{\mbox{$\x10^{#1}$}}
\newcommand{\x}{\times}
\newcommand{\bm}{\boldmath}
\newcommand{\ubm}{\unboldmath}
\newcommand{\beqn}{\begin{eqnarray}}
\newcommand{\eeqn}{\end{eqnarray}}
\newcommand{\beq}{\begin{eqnarray*}}
\newcommand{\eeq}{\end{eqnarray*}}
\newcommand{\pmo}{\mbox{$^{-1}$}}

\renewcommand{\v}[1]{\mbox{\bm$#1$\ubm}}

\newcommand{\BS}{Bulirsch-Stoer}
\newcommand{\eBS}{\mbox{$\varepsilon_{_{\rm BS}}$}}

\newcommand{\om}{\omega}
\newcommand{\Om}{\Omega}
\newcommand{\vpi}{\mbox{$\varpi$}}
\newcommand{\eps}{\mbox{$\epsilon$}}
\newcommand{\cn}{\mbox{$\v{\chi}_0$}}
\newcommand{\dn}{\mbox{$d_0$}}

\def\figdir{}
\def\ZBstd{ZB17d}

\shorttitle{Solar System Orbits}
\shortauthors{Zeebe}

\begin{document}

\title{Numerical Solutions for the orbital motion of the Solar System 
over the Past 100~Myr: Limits and new results {\Large $^\star$}}

\author{Richard E. Zeebe$^{1}$}
\affil{\vspace*{0.5cm}
     $^1$SOEST, 
     University of Hawaii at Manoa, 
     1000 Pope Road, MSB 629, Honolulu, HI 96822, USA. 
     email: zeebe@soest.hawaii.edu\\
%     Revised Version\\
     {\bf The Astronomical Journal, 154:193, 2017}\\ 
     October 29, 2017 \\
%    September 12, 2017 \\
     {\Large $^\star$} Numerical solutions are freely 
     available at: \myurl \\
     %\newpage \vspace*{-15ex}
     }

\renewcommand{\baselinestretch}{\blsA}\selectfont

\begin{abstract}
I report results from accurate numerical integrations 
of Solar System orbits over the past 100~Myr with the 
integrator package \hnb. The simulations used different 
integrator algorithms, step sizes, initial conditions,
and included effects from general relativity, different 
models of the Moon, the Sun's quadrupole moment, 
and up to sixteen asteroids. I also probed 
the potential effect of a hypothetical Planet~9, using 
one set of possible orbital elements. The most expensive 
integration (Bulirsch-Stoer) required 4~months wall-clock 
time with a maximum relative energy error $\lsim$3\e{-13}. 
The difference in Earth's eccentricity (\DeE) was used to 
track the difference between two solutions, considered to 
diverge at time $\tau$ when max$|\DeE|$ irreversibly crossed 
\sm10\% of mean \eE\ ($\sm0.028\x0.1$). The results indicate 
that finding a unique orbital solution is limited by initial 
conditions from current ephemerides and asteroid 
perturbations to \sm54~Myr. Bizarrely, the 4-month 
Bulirsch-Stoer integration and a symplectic integration
that required only 5~hours wall-clock time (12-day time step,
Moon as a simple quadrupole perturbation), agree to 
\sm63~Myr. Internally, such symplectic integrations 
are remarkably consistent even for large time steps, suggesting 
that the relationship between time step and $\tau$ is not a 
robust indicator for the absolute accuracy of symplectic 
integrations. The effect of a hypothetical Planet~9 on \DeE\ 
becomes discernible at \sm65~Myr. Using $\tau$ as 
a criterion, the current state-of-the-art solutions all 
differ from previously published results beyond \sm50~Myr.
I also conducted an eigenmode analysis, which provides some 
insight into the chaotic nature of the inner Solar System.
The current study provides new orbital solutions for 
applications in geological studies.
\end{abstract}


\renewcommand{\baselinestretch}{\bls}\selectfont
\section{Introduction}

The dynamical properties of the Solar System are of 
continuing interest to a number of fundamental, as well as 
applied research areas in a variety of scientific 
disciplines. For example, several studies have recently 
revisited the Solar System's dynamic stability 
on billion-year time scale using long-term
numerical integrations
\citep[e.g.,][]{batygin08,laskar09,zeebe15apjA,zeebe15apjB}.
Another area of active research concerns the application
of astronomical solutions for Earth's orbital parameters
in the geological sciences, with vital importance for astrochronology, 
cyclostratigraphy, and paleoclimatology
\citep[e.g.,][]{laskar11,westerhold12,mameyers17,zeebe17pa}.
Orbital solutions in geological applications are most 
heavily used over the past \sm100~Myr, which is the focus 
of the present study.
In particular, extending an astronomically-tuned 
geologic time scale further back in time would represent
a major advance in that field. Currently, the limit
to identifying a unique orbital solution is \sm50~Myr~BP,
as small differences in initial conditions or other
parameters cause astronomical solutions to diverge
around that time, due to the chaotic nature of the system. 
Hence one focus of the present study
will be to determine the divergence time between two
solutions (defined below). In addition to far-reaching
goals such as improvement of the geologic time scale, orbital 
solutions are key to a variety of specific applications, 
including resolving the precise 
timing of early Paleogene climate events of extreme
warmth (so-called hyperthermals), relative to orbital
forcing \citep[e.g.,][]{zachos08natNatB}.

It appears that the numerical solutions for Earth's 
eccentricity that have hitherto been used in paleoclimate 
studies were provided by only two different groups that
integrated the full Solar System equations over the past 
few 100~Myr \citep[e.g.,][]{varadi03,laskar04NatB,laskar11}.
Given one numerical realization of a Solar System model 
(i.e., via one code or integrator package),
various parameters determine the properties of the 
astronomical solution and usually limit its validity to 
a certain time period. Such limitations may be considered
internal and include limitations due to the underlying 
physics/physical model and numerics. The physics include,
for instance, initial coordinates and velocities of 
Solar System bodies, treatment of the Moon and asteroids, 
effects from general relativity,
the Sun's quadrupole moment, and the intrinsic dynamics 
of the system, e.g., its chaotic nature. 
Numerical issues include, for instance, the type of 
solver algorithm, numerical accuracy (e.g., time step), 
round-off errors, and choice of integrator coordinates
\citep{zeebe15apjA}. At present, internal limitations 
seem to restrict the validity of astronomical solutions 
to perhaps the past 50~Myr 
\citep{laskar11,laskar11ast}.
However, little is currently known about external 
limitations, that is, how different numerical realizations 
compare, say, between different investigator groups using 
different codes and integrator packages. Also, a one-to-one
comparison of orbital solutions obtained with 
fundamentally different algorithms such as 
\BS\ (BS) and symplectic integrators (yet otherwise 
identical setup) appears to be missing
\citep['symplectic integrators' here refers to $N$-body 
maps, e.g.,][]{wisdom91}. 

In this paper, I present new results from state-of-the-art 
Solar System integrations over the past 100~Myr to 
address the outstanding questions posed above.
I provide several new astronomical solutions for 
Earth's eccentricity for applications in geological studies.
\footnote{Numerical solutions are freely available 
at: \myurl}
I also investigated the
effect of a hypothetical Planet~9, though only for 
a single set of orbital elements as proposed in the 
literature. Furthermore, I performed an eigenmode analysis
to gain some insight into the chaotic behavior of the 
system. Finally, I discuss the main factors that currently
limit identification of a unique orbital solution beyond 
\sm50~Myr.

\section{Methods} \label{SecMeth}

The integrations were performed with the integrator 
package \hnb\ \citep{rauch02} using the \BS\
integrator with relative accuracy \eBS\ and the 
symplectic integrator \citep{wisdom91}
with time step \Dt\ 
(Table~\ref{TabRuns}). Relativistic 
corrections \citep{einstein16} are critical 
\citep{varadi03,laskar04NatB,zeebe15apjA}
and are available in \hnb\ as Post-Newtonian effects due 
to the dominant mass. Hence all simulations presented 
here include contributions from general relativity 
(GR). Also, all symplectic integrations were carried
out using Jacobi coordinates \citep{wisdom91}, rather than 
heliocentric coordinates \citep{zeebe15apjA}. 
In this study, wall-clock times refer to {\tt 
\hnb-v1.0.10}, double precision on 64-bit Linux 
machines with Intel i7-3770 3.40 GHz cores.

\subsection{Treatment of the Moon}

The Moon was included as a separate object
(BS and symplectic option \ems), or the Earth-Moon system
was modeled as a gravitational quadrupole (symplectic 
option \lun, see Table~\ref{TabRuns}) 
\citep{quinn91,varadi03,rauch02}. The \ems\ option
includes a symplectic, self-consistent sub-integration of 
the Earth-Moon-Sun system; the remaining 
Solar System bodies treat the Earth-Moon barycenter 
as a single object \citep{rauch02}. The \lun\ 
option considers the Moon's influence on the net motion 
of the Earth-Moon barycenter via a mean
quadrupole potential with a correction factor
$f_{\tt lun} = 0.8525$ \citep{quinn91,varadi03,rauch02}.
The effect of tidal dissipation in the Earth-Moon 
system was tested in the solution ZB17k following 
\citet{quinn91}, 
using a constant secular change in semimajor axis 
close to the modern value of $\dot{a} \simeq 3.8$~cm~y\pmo\ 
over the past 100~Myr (Table~\ref{TabRuns}). 
However, this effect was not included in other runs 
for two reasons. First, the modern situation is a 
poor analogue for the past, where $\dot{a}$ was
most likely smaller but has large uncertainties 
\citep[e.g.,][]{green17}. Second, tidal dissipation 
in the Earth-Moon system turned out to have a 
minor effect on the results compared to other parameters 
(see Section~\ref{SecRes}).

\def\tx{-.5ex}
%---------------------------------------------------------%
%-------------------- TABLE ------------------------------%
%---------------------------------------------------------%
\begin{table*}[t]
\caption{Summary of numerical solutions. \label{TabRuns}}
\vspace*{5mm}
\hspace*{-5mm}
\begin{tabular}{llllllr}
\tableline\tableline
Solution & Algorithm   & \eBS\ or $\Dt$ & Moon     & \v{x_0,v_0} & $J_2$-rot  & Asteroids \\
\hline     
ZB17a$^a$& BS$^b$      & 1\e{-15}         & separate & DE431       & BG05$^c$   & 10      \\ [\tx]
ZB17b    & 4th Sympl$^{TT}$ & 0.375~d     & ems      & DE431       & BG05       & 10      \\ [\tx]
ZB17c    & 4th Sympl$^{TT}$ & 0.375~d     & lun      & DE431       & BG05       & 10      \\ [1.5ex]
\cline{3-5}
ZB17d    & 2nd Sympl        & 2.0~d       & lun      & DE431       & BG05       & 10      \\ [\tx]
ZB17e    & 2nd Sympl        & 2.0~d       & lun      & INPOP13c    & BG05       & 10      \\ [\tx]
ZB17f    & 2nd Sympl        & 2.0~d       & lun      & DE431       & HCI$^d$    & 10      \\ [\tx]
ZB17g    & 2nd Sympl        & 2.0~d       & lun      & DE431       & BG05       & big 3   \\ [\tx]
ZB17h    & 2nd Sympl        & 2.0~d       & lun      & DE431       & BG05       & 8       \\ [\tx]
ZB17i    & 2nd Sympl        & 2.0~d       & lun      & DE431       & BG05       & 13      \\ [\tx]
ZB17j    & 2nd Sympl        & 2.0~d       & lun      & DE431       & BG05       & 16      \\ [\tx]
ZB17k    & 2nd Sympl        & 2.0~d       & lun,TD$^e$ & DE431     & BG05       & 10      \\ [\tx]
ZB17p$^p$& 2nd Sympl        & 2.0~d       & lun      & DE431       & BG05       & 10      \\ [1.5ex]
\cline{3-5}
s405$^v$ & 2nd Sympl        & 2.0~d       & lun      & DE405       & $-$        & $-$     \\ [\tx]
sL11$^l$ & 2nd Sympl        & 2.0~d       & lun      & INPOP10a    & BG05       & 5       \\
\tableline
\end{tabular}

\noindent {\small
$^a$ZB = {\bf Z}eebe-HN{\bf B}ody.                           \\ [\tx]
$^b$BS = \BS.                                                \\ [\tx]
$^c$BG05 = \citet{beckgiles05}, see Section~\ref{SecSolRot}. \\ [\tx]
${TT}$ = TipToe \hnb\ option (extra careful Kepler drifts).  \\ [\tx]
$^d$HCI = Heliocentric Inertial SPICE frame.                 \\ [\tx]
$^e$TD = Includes tidal dissipation in Earth-Moon system.    \\ [\tx]
$^p$Planet 9.                            \\ [\tx]
$^v$Test run for comparison with \citet{varadi03}'s R7.      \\ [\tx] 
$^l$Test run for comparison with \citet{laskar11ast}.
}
\end{table*}
%---------------------------------------------------------%

\subsection{Initial conditions and asteroids}

Different sets of initial conditions for the positions 
and velocities $\cn := (\v{x}_0, \v{v}_0)$ of the planets
and Pluto were employed based on the ephemerides DE431
\citep[released in 2013,][]{folkner14}, DE405
\citep{standish98}, INPOP13c 
\citep{fienga14}, and INPOP10a \citep{fienga11}.
For example, DE431 covers years $-$13,200 to +17,191; 
INPOP13c is available for J2000 $\pm$1,000 years.
The ephemerides are generated by fitting numerically 
integrated orbits to observations. Ephemeris integrations
usually use models of a high degree of completeness and 
are computationally expensive. One option to obtain initial 
conditions from ephemerides is by fitting the long-term 
integration models to 
ephemerides over a certain time interval \citep[e.g.,][]{laskar11}.
It appears that this would modify the initial 
conditions so as to compensate for the differences between 
the less complete long-term models and the more sophisticated 
ephemeris models. Clearly, this approach will lead to 
improved agreement between the two models over the fit interval.
However, will such initial conditions guarantee the most accurate 
results in the long-term integration
across the time intervals not covered by ephemerides?
For example, DE431 includes 343 asteroids in the dynamical
model, whereas long-term models may include only a few 
(up to sixteen here) or none at all. Thus, the difference
between the long-term- and ephemeris models is of dynamical
nature (mutual interactions between Solar System bodies)
that persists throughout the entire integration. In
contrast, initial conditions affect the positions and 
velocities of only those Solar System bodies included in 
the integration 
and at one particular point in time. Hence the two issues
(dynamical model vs.\ initial conditions) relate to 
different aspects of the integration, which do not 
necessarily need to cancel each other out in the long run.
Another approach is to directly adopt the initial conditions 
from ephemerides at an epoch that is covered
by modern observations \citep[e.g.,][]{varadi03}. 
The latter approach was used here.

For DE431 and DE405
(\url{naif.jpl.nasa.gov/pub/naif/generic_kernels/spk/planets}),
\cn\ was generated using the SPICE toolkit for Matlab 
(\url{naif.jpl.nasa.gov/naif/toolkit.html}).
For INPOP (\url{www.imcce.fr/inpop}), \cn\
was generated using the calceph library in C
(\url{www.imcce.fr/inpop/calceph}). Coordinates
were obtained at JD2451545.0 (01 Jan 2000, 12:00 
TDB = J2000.0) in certain 
inertial reference frames and subsequently rotated 
\citep[cf.,][]{souami12} if applicable 
(see Section~\ref{SecSolRot}). In the following,
ICRF (International Celestial Reference Frame, $\equiv$ 
J2000 in SPICE) refers to Earth's mean equator
and dynamical equinox of J2000.0; ECLIPJ2000 refers
to ecliptic coordinates based on the J2000 frame
(\url{naif.jpl.nasa.gov/pub/naif/toolkit_docs/C/req/frames.html}).

Initial conditions for the asteroids were generated at 
\url{ssd.jpl.nasa.gov/x/spk.html}. All asteroids
were treated as heavyweight particles (HWPs) in \hnb,
i.e., subject to the same, full interactions as the
planets and Pluto. The runs labeled "big~3" (Table~\ref{TabRuns}) 
include the asteroids Vesta, Ceres, and Pallas. 
Additional asteroids considered in other 
simulations were included in the order given in 
Table~\ref{TabAst}. The test solution "s405" attempts to 
replicate simulation R7 \footnote{Also available at: 
\myurl} of \citet{varadi03}, who did not 
include asteroids. The solution "sL11" represents a test 
run for comparison with \citet{laskar11ast}, including
5~asteroids and initial conditions based on INPOP10a
\citep{fienga11}.

\def\tx{-1ex}
%---------------------------------------------------------%
%-------------------- TABLE ASTEROIDS --------------------%
%---------------------------------------------------------%
\begin{table}[h]
\caption{Order of asteroids included in different simulations.$^a$
         \label{TabAst}}
\vspace*{5mm}
\hspace*{-5mm}
\begin{tabular}{rlc}
\tableline\tableline
\#       & Name       & Mass$^b$ \\
\tableline   
 1 & Vesta      & 1.30E-10 \\ [\tx]
 2 & Ceres      & 4.73E-10 \\ [\tx]
 3 & Pallas     & 1.05E-10 \\ [\tx]
 4 & Iris       & 7.22E-12 \\ [\tx]
 5 & Bamberga   & 4.69E-12 \\ [\tx]
 6 & Hygiea     & 4.18E-11 \\ [\tx]
 7 & Euphrosyne & 2.14E-11 \\ [\tx]
 8 & Interamnia & 1.78E-11 \\ [\tx]
 9 & Davida     & 1.76E-11 \\ [\tx]
10 & Eunomia    & 1.58E-11 \\ [\tx]
11 & Juno       & 1.22E-11 \\ [\tx]
12 & Psyche     & 1.15E-11 \\ [\tx]
13 & Cybele     & 1.07E-11 \\ [\tx]
14 & Thisbe     & 8.71E-12 \\ [\tx]
15 & Doris      & 8.55E-12 \\ [\tx]
16 & Europa     & 8.37E-12 \\
\tableline
\end{tabular}

\noindent {\small
$^a$Simulations with $N$~asteroids (see Table~\ref{TabRuns})
	include objects \#1 to \#$N$.         \\[0ex]
$^b$In solar masses \citep{folkner14}.        \\[0ex]
}
\end{table}
%---------------------------------------------------------%

\subsection{Planet~9}

The perturbation of a hypothetical Planet~9 (P9) on 
Earth's eccentricity was examined using one set of orbital 
elements as proposed in the literature. Note that P9's 
existence and hence its orbit is entirely speculative at 
this point \citep{trujillo04,brownbatygin16,batyginbrown16,
fienga16,malhotra16,holman16,shankman17,millholland17}.
Thus, a large array of orbits is possible and the purpose
of the current simulations is merely to test on which
time scale a distant perturber would cause a noticeable difference 
in Solar System trajectories over 100~Myr (see 
Section~\ref{SecRes}). P9's assumed elements/mass were: semimajor 
axis $a = 654$~AU, eccentricity $e = 0.45$, inclination
$I = 30\deg$, longitude of ascending node $\Om =
50\deg$, argument of perihelion $\om = 150\deg$,
mean anomaly $M = 180\deg$, and mass $m = 10 \x \mE$
\citep{millholland17}.

\subsection{Solar Rotation Axis and Quadrupole Moment $J_2$}
\label{SecSolRot}

Recent studies have converged on a value for the 
solar quadrupole moment $J_2$ of \sm$2.2\e{-7}$
\citep[e.g.,][]{pijpers98,mecheri04,fienga15j2,pitjeva14,
park17j2}, which was used here throughout.
The solar quadrupole moment is directed 
along the solar rotation/symmetry axis, which is 
about $6\deg$ and $7\deg$ offset from the invariable 
plane and ECLIPJ2000, respectively
\citep{carrington1863,giles00,fraenz02,beckgiles05,
baileybatygin16}. The initial (Cartesian) coordinates
were hence rotated to account for this offset \citep{souami12,
fraenz02}. By default, the quadrupole in \hnb\ is 
directed along the z-axis, which was taken
as the solar rotation axis. Traditional Carrington
elements for inclination and longitude of ascending 
node of the solar equator relative to ECLIPJ2000 are 
$i_\Sun = 7.25\deg$ and $\Om_\Sun = 75.76\deg$
at J2000.0, respectively \citep{fraenz02}. 
However, most runs performed here use
more recent values of $i_\Sun = 7.155\deg$ and 
$\Om_\Sun = 75.594\deg$ \citep{beckgiles05} at J2000.0
(labeled BG05 in Table~\ref{TabRuns}). 
In principle, the coordinate transformation described 
above is equivalent to expressing the coordinates 
in the Heliocentric Inertial (HCI) frame in SPICE, 
except the latter uses the declination $\delta_\Sun = 
63.87\deg$ and the right ascension $\alpha_\Sun = 
286.13\deg$ of the solar rotation axis (solution 
ZB17f, option $J_2$-rot = HCI, Table~\ref{TabRuns}). 
The s405 setup ($J_2 = 0$) again follows \citet{varadi03}, 
who, it appears, did not consider $J_2$. 

\section{Results} \label{SecRes}

\begin{figure}[t]
\def\sc{0.62}
\begin{center}
\includegraphics[scale=\sc]
{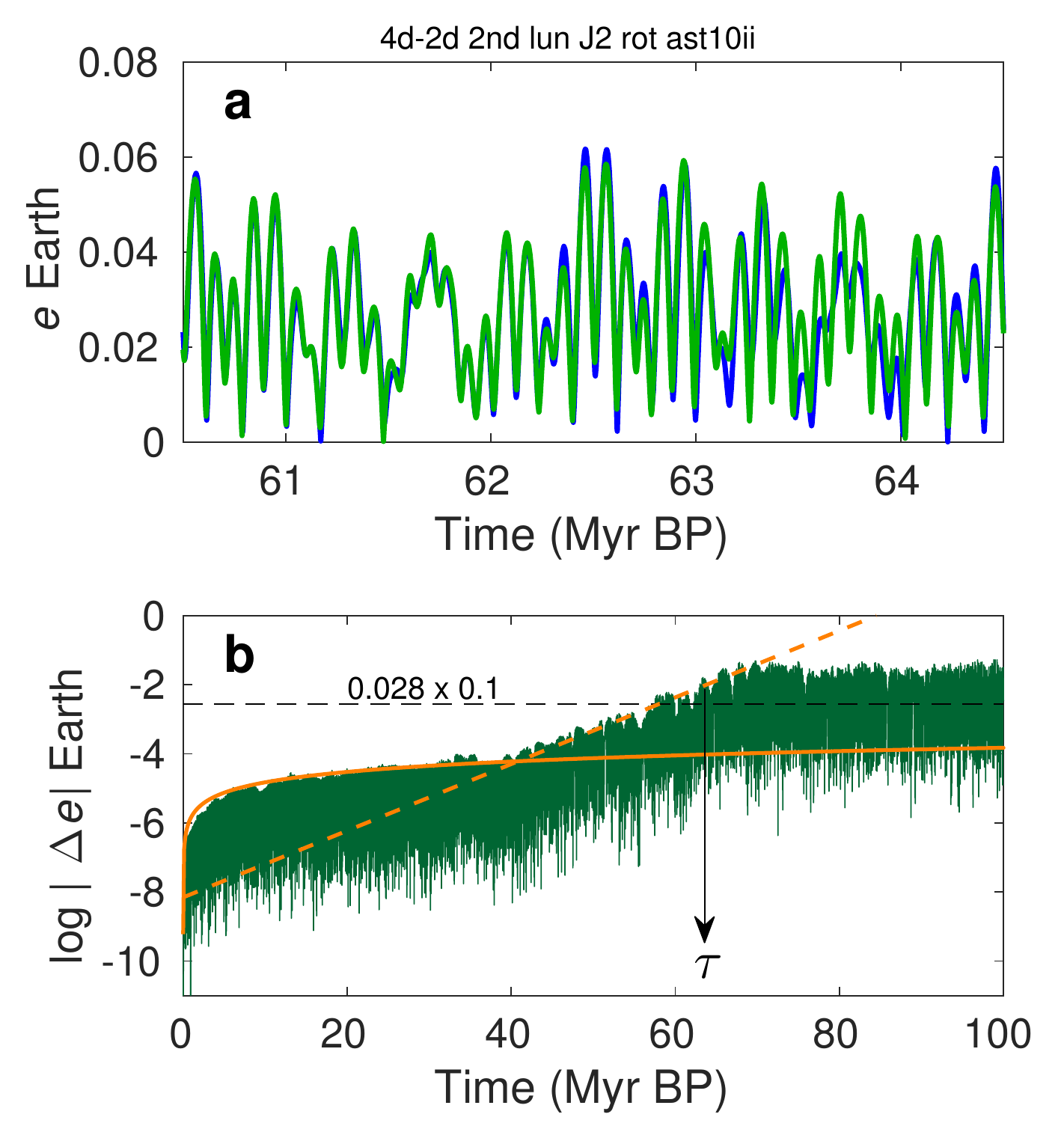}
\end{center}
\caption{\fns (a) Earth's orbital eccentricity for the base 
settings of \ZBstd\ (Table~\ref{TabRuns}) using the \hnb\ 
2nd-order symplectic integrator at $\Dt = 4$ (green) vs.\ 2~days
(blue). (b) Corresponding difference in Earth's eccentricity, 
$|\DeE|$, over the past 100~Myr. The arrow indicates
the divergence time $\tau$, when max$|\DeE|$ irreversibly 
crosses \sm10\% of mean \eE\ ($\sm0.028\x0.1$, dashed line).
Orange curves: simple fit functions with linear growth
in $|\DeE|$ (solid) and exponential growth in $|\DeE|$ 
(dashed, linear on log-$y$ scale) with a Lyapunov time of
4.5~Myr (see text).
}
\label{FigZB4d-2d}
\end{figure}

In the following, the difference between two orbital
solutions will be tracked by the divergence 
time $\tau$, i.e., the time when the difference 
in Earth's eccentricity 
(\DeE) irreversibly crosses \sm10\% of mean \eE\ 
($\sm0.028\x0.1$, Fig.~\ref{FigZB4d-2d}). The 
divergence time $\tau$
as used here should not be confused with the Lyapunov time,
which is the time scale of exponential divergence
of trajectories and is only \sm5 Myr for the inner planets
\citep{laskar90,varadi03,batygin08,zeebe15apjA}.
For the solutions discussed here, the divergence
of trajectories is ultimately dominated by exponential 
growth ($t \gsim 40$~Myr~BP), which is indicative of chaotic 
behavior (Fig.~\ref{FigZB4d-2d}). Thus, $\tau$ is largely 
controlled by the Lyapunov time, though the two are of 
course different quantities. Integration errors usually 
grow polynomially and typically dominate for $t \lsim 
40$~Myr~BP (see Fig.~\ref{FigZB4d-2d} and e.g., 
\citet{varadi03}).

%-----------------------------------------------------------%
%--------------------  FIGURE ------------------------------%
%-----------------------------------------------------------%
\renewcommand{\baselinestretch}{\blsC}\selectfont
\renewcommand{\baselinestretch}{\bls}\selectfont
%-----------------------------------------------------------%

\subsection{Numerical algorithm and step size} \label{SecNum}

\begin{figure}[t]
\def\sc{0.62}
\begin{center}
\includegraphics[scale=\sc] {\figdir 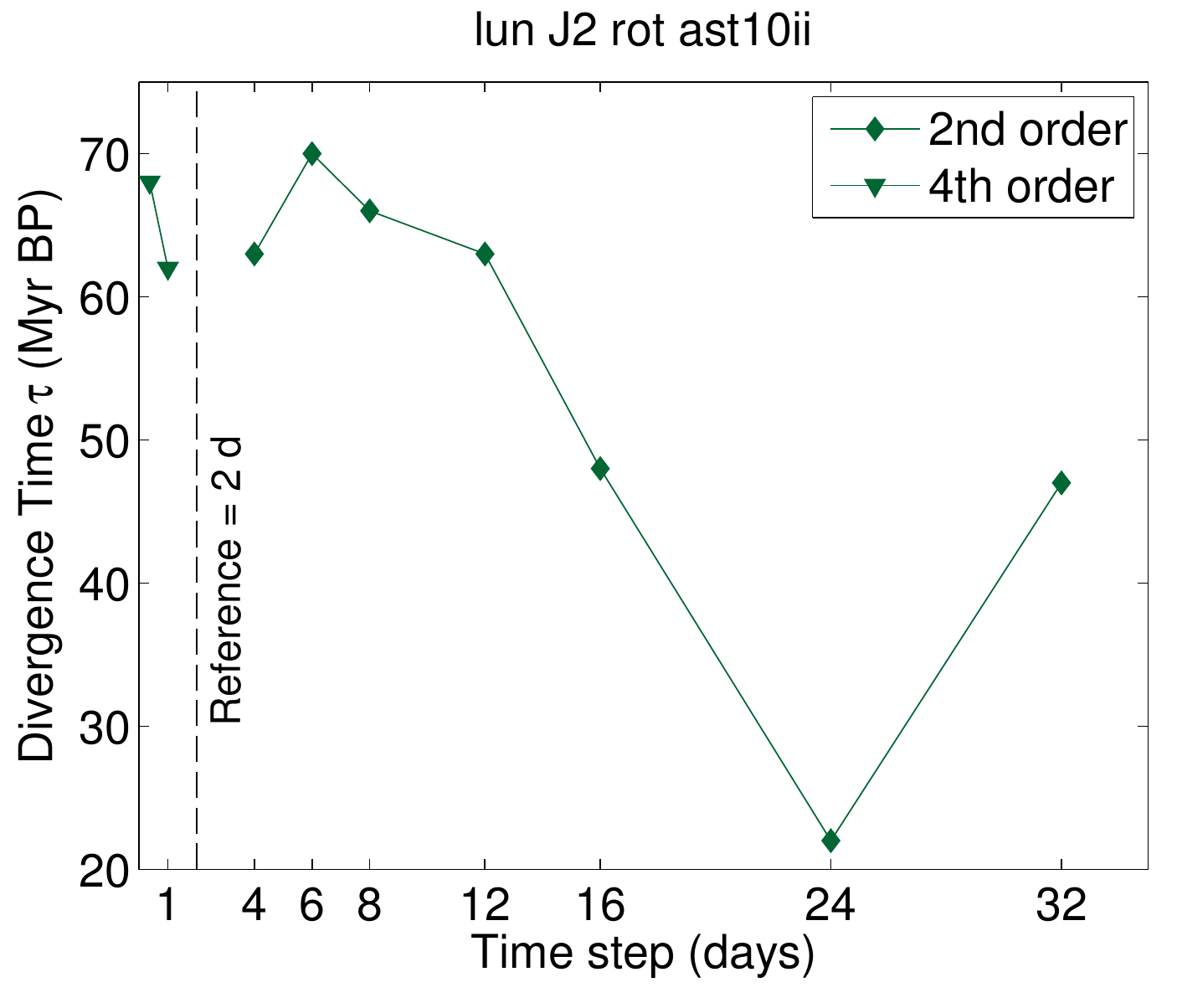}
\end{center}
\caption{\small Divergence times for solutions based
on the settings of \ZBstd\ (\lun\ option,
Table~\ref{TabRuns}) using 
the \hnb\ 2nd- and 4th-order symplectic integrator at 
various time steps, relative to the reference solution 
obtained at $\Dt = 2$~days (2nd-order, vertical
dashed line). The 4th-order time steps are 0.375~d 
and 1~d, respectively.
}
\label{FigTauDt}
\end{figure}

A given numerical algorithm is often evaluated by
varying its accuracy or step size (\Dt), while
keeping all other parameters constant. For instance,
for the base settings of \ZBstd\ (Table~\ref{TabRuns}),
\hnb's 2nd-order symplectic integrator gives 
$\tau \simeq 63$~Myr at $\Dt = 4$ vs.\ 2~days
(Fig.~\ref{FigZB4d-2d}). One might hence assume that
$\tau$ reflects the time span of the validity of 
the symplectic solution when using progressively 
smaller step sizes and that $\tau$ would drop off rapidly 
for larger \Dt. However, this is not necessarily the 
case, as test runs with the \ZBstd\ setup (\lun\
option) and different 
\Dt\ show (Fig.~\ref{FigTauDt}). Using $\Dt = 2$~d as the 
reference case, the symplectic runs with larger time steps
of 4, 6, 8, and 12~days all show \textbf{\textit{ 
larger}} $\tau$'s than with a smaller time step 
of 1~day, which appears counterintuitive.
One might expect \textbf{\textit{smaller}} $\tau$'s
at larger time steps (supposedly less accurate). Only for
$\Dt\ \gsim 12$~d, $\tau$ starts to fall off. Surprisingly, 
even for an absurdly large time step of 32~days, the 
symplectic solution only diverges from the reference case 
with 2-day time step at \sm47~Myr (Fig.~\ref{FigTauDt},
all \lun\ option).
These results suggest that the relationship between
divergence time and time step is not not a robust 
indicator for the accuracy of symplectic solutions.

%-----------------------------------------------------------%
%--------------------  FIGURE ------------------------------%
%-----------------------------------------------------------%
\renewcommand{\baselinestretch}{\blsC}\selectfont
\renewcommand{\baselinestretch}{\bls}\selectfont
%-----------------------------------------------------------%

\begin{figure*}[t]
\def\sc{0.60}
\begin{center}
\includegraphics[scale=\sc] {\figdir 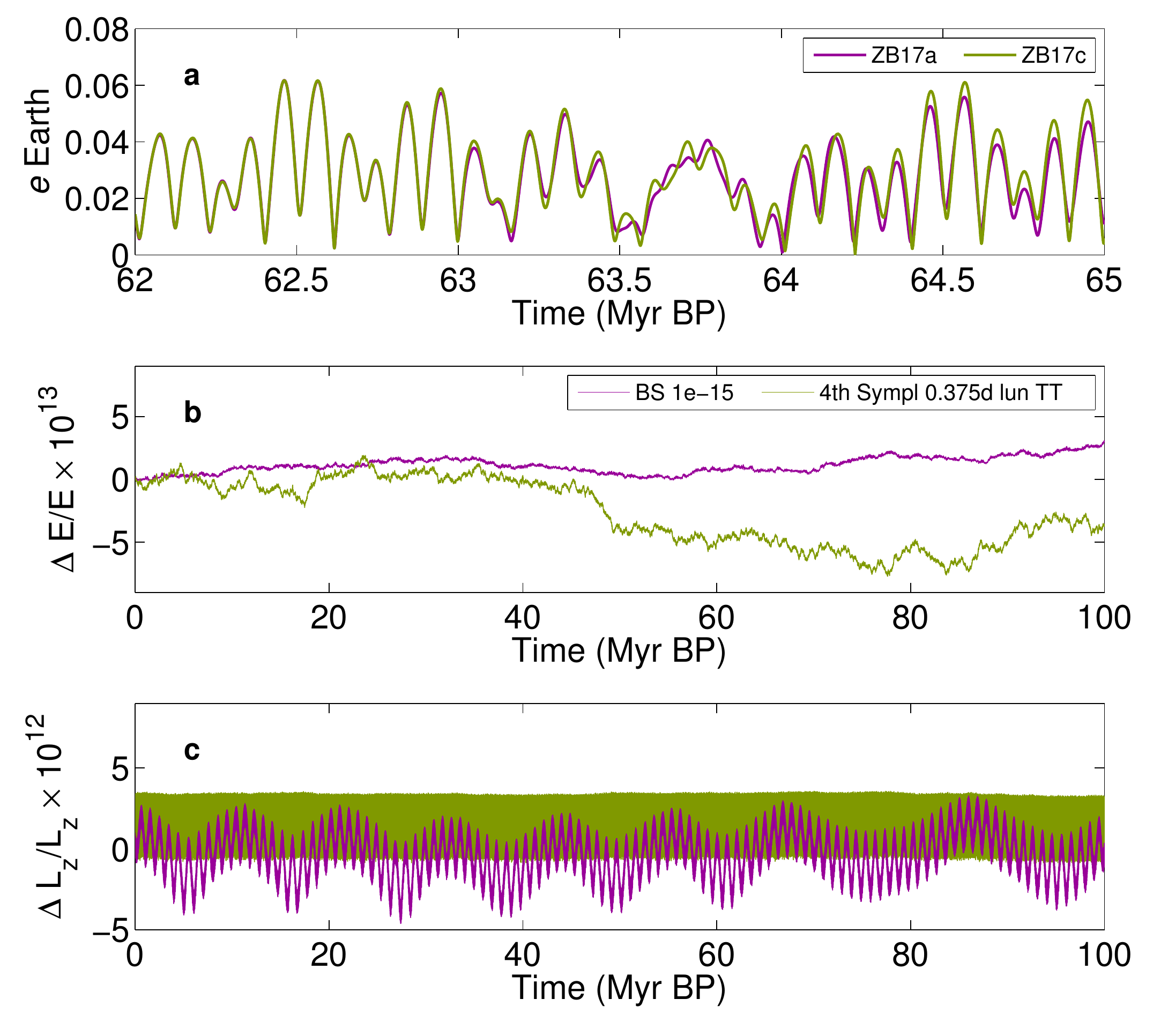}
\end{center}
\caption{\small
(a) Solutions ZB17a (\BS, purple) and 
ZB17c (symplectic, green).
Maximum relative error in (b) total energy 
$|\D E/E| = |(E(t)-E_0)/E_0|$ and (c) in $z$-component of 
angular momentum ($|\D L_z/L_z|$). 
Note that steps in the symplectic energy due
to close encounters of asteroids have been removed.
{TT} = TipToe \hnb\ option (extra careful Kepler drifts).
}
\label{FigdEEdLL}
\end{figure*}

Moreover, integrations with different
algorithms (symplectic vs.\ \BS\ at small 
$\Dt = 0.375$~d and relative accuracy $\eBS = 1\e{-15}$,
respectively)
diverge at \sm63~Myr (Fig.~\ref{FigdEEdLL}).
This is earlier than some solutions obtained with just a 
single algorithm, e.g., the symplectic integrations for 
$\Dt\ = 2$ vs.\ 6~days discussed above ($\tau \simeq 
70$~Myr). Hence, while symplectic integrations 
with \lun\ option appear internally remarkably 
consistent at different time steps, 
this does not necessarily imply that symplectic solutions 
are reliable up to $\tau$, as the
comparison with a different algorithm shows. At this 
stage, it remains inconclusive which numerical algorithm 
provides more accurate solutions for the problem at hand.

%-----------------------------------------------------------%
%--------------------  FIGURE ------------------------------%
%-----------------------------------------------------------%
\renewcommand{\baselinestretch}{\blsC}\selectfont
\renewcommand{\baselinestretch}{\bls}\selectfont
%-----------------------------------------------------------%

For example, the most expensive simulation of the present
study, i.e., the 4-month-long BS integration (ZB17a) at 
relative accuracy $\eBS = 1\e{-15}$ shows excellent 
conservation of energy and angular momentum (max$|\D E/E| 
\ \lsim \ 3\e{-13}$, max$|\D L_z/L_z| \ \lsim \ 5\e{-12}$, 
Fig.~\ref{FigdEEdLL}). This is an important consideration 
for BS-integrations, which 
suffer from significant $E$- and $L$-drifts if the time 
step is too large. However, $E$ and $L_z$ are nearly 
equally well preserved in the 4th-order symplectic 
integration ZB17c with the smallest 
time step used ($\Dt = 0.375$~d, Fig.~\ref{FigdEEdLL}). 
Thus, energy- and angular momentum conservation do not 
provide a criterion here for selecting one solution/algorithm
over the other.

Establishing quality criteria just among symplectic 
integrators based on energy properties also appears 
elusive. Usually,
the long-term energy drift and fluctuations around the 
short-term mean increase with the symplectic step size. For
example, the symplectic $\D t = 12$~d-run has \sm100$\x$ and 
\sm10$\x$ larger standard deviation and energy drift,
respectively, than the $\D t = 0.375$~d-run 
(ZB17c, Fig.~\ref{FigTauDt}). One might therefore 
assume the ZB17c solution to be more accurate. However, 
both symplectic solutions (0.375~d and 12~d time step) 
diverge from ZB17a at \sm63~Myr, providing
no conclusion about accuracy. Furthermore, this 
leads to the bizarre conclusion that the 4-month 
Bulirsch-Stoer integration and the 5-hour (wall-clock 
time) symplectic integration with \lun\ option
at $\D t = 12$~d 
give essentially the same results to 
\sm63~Myr. Thus, full Solar System integrations 
for, e.g., parameter studies over $\lsim$60~Myr may be 
completed within a few hours, rather than months.

\def\s{\hspace*{2ex}}
\def\tx{2ex}
%---------------------------------------------------------%
%-------------------- TABLE ------------------------------%
%---------------------------------------------------------%
\begin{table*}[t]
\caption{Selected, approximate divergence times ($\tau$ in Myr 
BP) for pairs of solutions $\cS_{ij}$. \label{TabTau}}
\vspace*{5mm}
\hspace*{-5mm}
\begin{tabular}{lccccccccccccccc}
\tableline\tableline
$\cS^s$  & a  & b  & c\s& d  & e  & f  & g  & h  & i  & j  & k  & p\s& V03$^v$ & La11$^l$ & La04$^m$ \\
\hline     
a        &    & 54 & 63 & 63 &    &    &    &    &    &    &    &    & 41  & 50  & 41   \\ [\tx]
b        &    &    & 54 & 54 &    &    &    &    &    &    &    &    & 41  & 50  & 41   \\ [\tx]
c        &    &    &    & 68 &    &    &    &    &    &    &    &    & 41  & 50  & 41   \\ [0ex]
d        &    &    &    &    & 54 & 63 & 48 & 56 & 54 & 54 & 63 & 65 &     &     &      \\ [\tx]
s405     &    &    &    &    &    &    &    &    &    &    &    &    & 54  &     &      \\ [\tx]
sL11     &    &    &    &    &    &    &    &    &    &    &    &    &     &  47 &      \\
\tableline
\end{tabular}

\noindent {\small \\
$^s$ Solutions, see Table~\ref{TabRuns}: a = BSe-15, b = 0.375d.ems, 
c = 0.375d.lun, d = 2nd2d.lun, e = inpop13c, f = hci, g = big3,
h = ast8, i = ast13, j = ast16,
k = tidal-dissipation, p = Planet~9. \\
$^v$ \citet{varadi03}, run R7.       \\
$^l$ \citet{laskar11ast}.  $\tau$(ZB17a-La10$x$) = [41 50 50 50], 
where $x = $\ a,b,c,d.               \\ 
$^m$ \citet{laskar04NatB}.           \\[-2ex] 
\rule{4cm}{0.4pt}
} 
\end{table*}
%---------------------------------------------------------%

\subsection{Test against previous solutions} \label{SecTest}

Further insight into the behavior of numerical orbital 
solutions may be gained by testing whether previous simulations 
can be reproduced when the same assumptions for the 
underlying physical model of the Solar System are used. 
For example, \citet{varadi03} (V03 for short) used a 
St{\"o}rmer scheme to integrate
the orbits of the major planets over the past 207~Myr
(their simulation R7), including GR corrections. V03's
initial conditions were taken from DE405 \citep{standish98},
while the Moon's influence on the net motion of
the Earth-Moon barycenter was modeled in R7 via a mean
quadrupole potential with a correction factor
$f_{\tt lun} = 0.8525$. Based on the information
provided in V03, effects of $J_2$ and 
asteroids were not included in the computations.

\begin{figure*}[t]
\def\sc{0.65}
\begin{center}
\includegraphics[scale=\sc]{\figdir 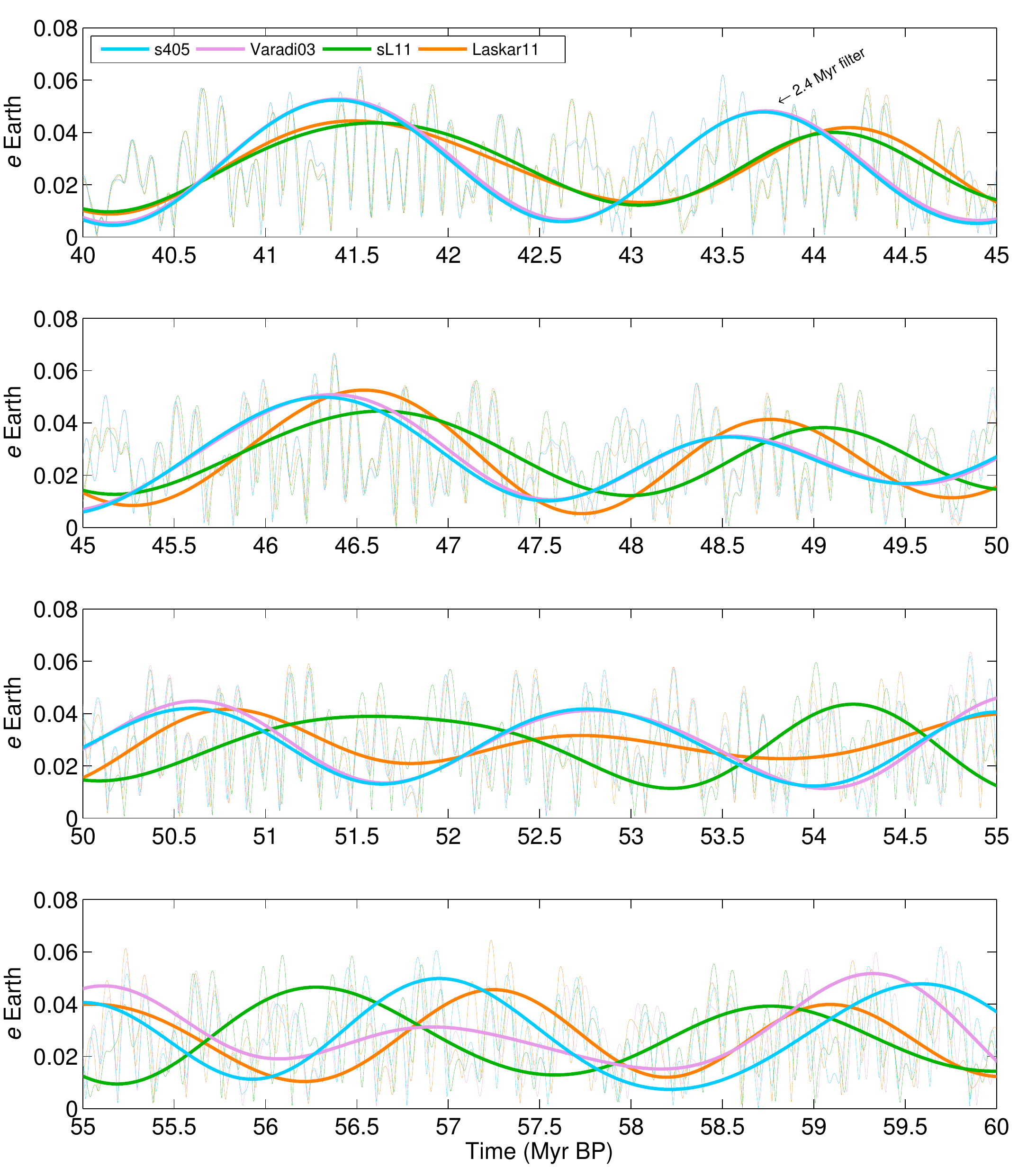}
\end{center}
\caption{\small Comparison of Earth's computed 
eccentricity (thin lines) of previously published 
and current test solutions. For labels and features of 
different solutions, see text and Tables~\ref{TabRuns} 
and~\ref{TabTau}. Also, a \sm2.4-Myr Gaussian filter
was applied to all solutions to aid in identifying 
differences in the low-frequency range around $g_4-g_3$ 
(thick lines, arbitrary scaling). Gaussian filter parameters
are: $\sigma = df / \sqrt{2\ln(2)}$ with half-width $df = 
0.5 \cdot f$; i.e., bandwidth $f \pm50$\% where 
$f = 0.4209$~Myr\pmo.
}
\label{FigV03}
\end{figure*}

The current test simulation s405 (Table~\ref{TabRuns})
uses the same physical model as V03 but was integrated
using \hnb's 2nd-order symplectic integrator with 
$\Dt = 2$~days. Solutions s405 and V03-R7 diverge at 
$\tau \simeq 54$~Myr
(Table~\ref{TabTau}, Fig.~\ref{FigV03}). This results is 
encouraging in terms of reproducibility, given that
different integrator algorithms were used and the fact that 
several other pairs of solutions diverge earlier 
(Table~\ref{TabTau}).

%-----------------------------------------------------------%
%--------------------  FIGURE ------------------------------%
%-----------------------------------------------------------%
\renewcommand{\baselinestretch}{\blsC}\selectfont
\renewcommand{\baselinestretch}{\bls}\selectfont
%-----------------------------------------------------------%

\begin{figure*}[t]
\def\sc{0.65}
\begin{center}
\includegraphics[scale=\sc]{\figdir 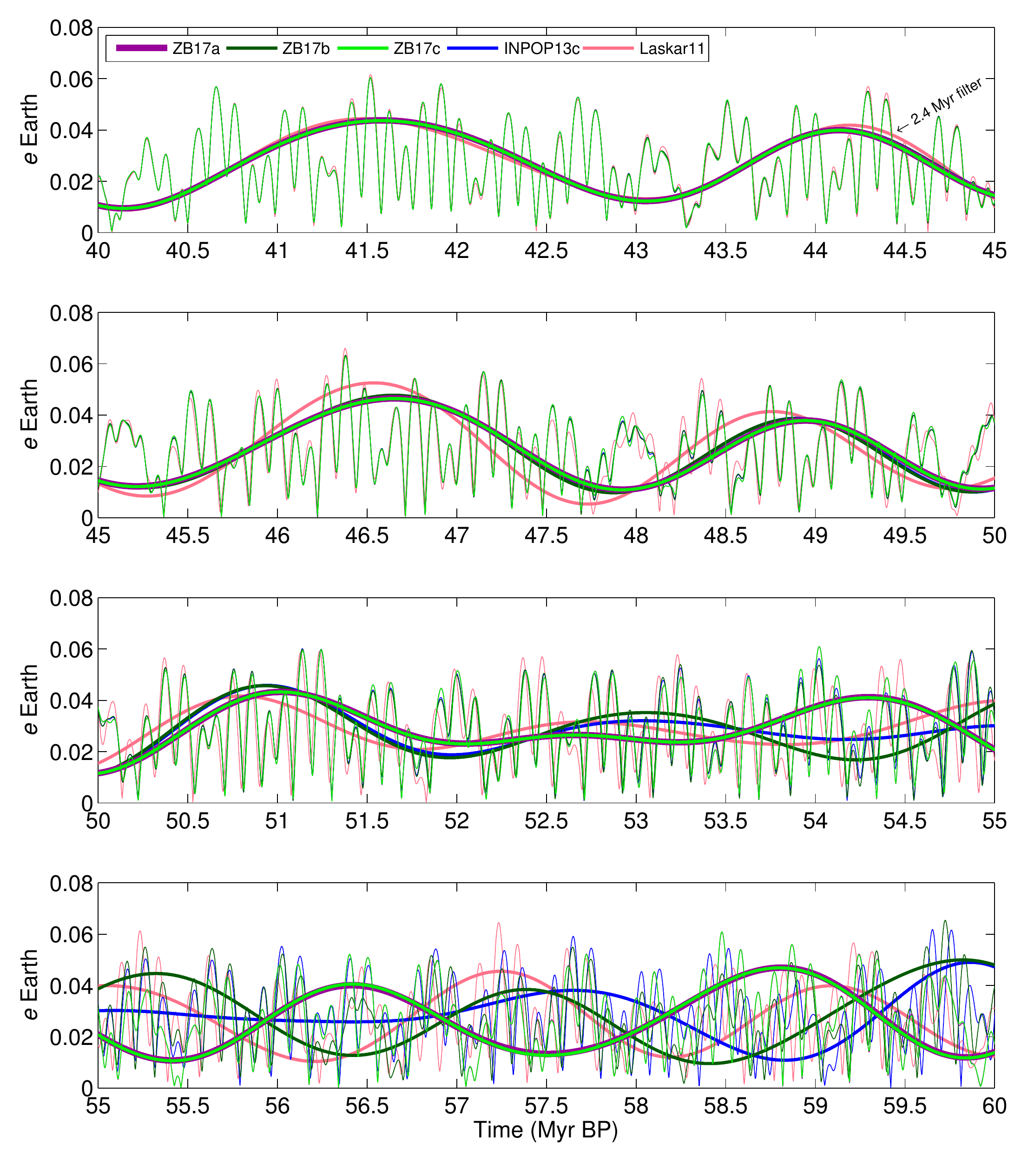}
\end{center}
\caption{\small 
Earth's computed eccentricity (thin lines) of 
selected current solutions and La11. For details, 
see text and Tables~\ref{TabRuns} and~\ref{TabTau}.
Gaussian filter (thick lines) as in Fig.~\ref{FigV03}
with arbitrary scaling. 
Solutions ZB17a and ZB17c agree to 63~Myr,
thus ZB17a's line width (purple, filtered) was 
increased for visibility. Note that for, e.g., La11-ZB17c 
(Table~\ref{TabTau}), the definition of $\tau$ 
(cf., Fig.~\ref{FigZB4d-2d})
gives a divergence time of \sm50~Myr, while the filtered 
La11 curve could connote a slightly extended agreement. 
However, La11 and ZB17c are irreversibly out-of-phase 
beyond \sm50~Myr, which causes max$|\DeE|$ to cross 
the threshold.
}
\label{FigZB17}
\end{figure*}

Using the Solar System model and symplectic integrator 
from the long-term solution La10 \citep{laskar11}, 
\citet{laskar11ast} provided orbital solutions over the 
past 100~Myr (La11 for short). Initial conditions were
based on INPOP10a \citep{fienga11,westerhold12} and 
the Moon plus 5~asteroids (Ceres, Pallas, Vesta, Iris, 
and Bamberga) were included as separate, full-interacting 
objects. The present test solution sL11
uses a similar setup but uses the \lun\ option and 
\hnb's 2nd-order, 2-day step integrator 
(Table~\ref{TabRuns}). The solar rotation axis and 
quadrupole moment were included as described in
Section~\ref{SecSolRot} using the BG05 option.
The solutions sL11 and La11 diverge at \sm47~Myr 
(Table~\ref{TabTau}, Fig.~\ref{FigV03}) and hence 
notably earlier than for the \citet{varadi03} test 
case. The reason could be differences in setup
and integrator, as mentioned above.
Interestingly, when 10~asteroids are included, the 
solution ZB17b, for instance, stays closer to 
La11 than sL11 ($\tau \simeq 50$~Myr, 
Table~\ref{TabTau}, Fig.~\ref{FigZB17}).

%-----------------------------------------------------------%
%--------------------  FIGURE ------------------------------%
%-----------------------------------------------------------%
\renewcommand{\baselinestretch}{\blsC}\selectfont
\renewcommand{\baselinestretch}{\bls}\selectfont
%-----------------------------------------------------------%

\subsection{Orbital solutions ZB17a,b,c} \label{SecABC}

The orbital solutions ZB17a,b,c are based on the most 
expensive integrations presented here (Table~\ref{TabRuns}, 
Fig.~\ref{FigZB17}). Remarkably, while the BS option 
(ZB17a) and the symplectic \lun\ option (ZB17c) diverge 
at \sm63~Myr, the \ems\ option separates much 
earlier, \sm54~Myr (Table~\ref{TabTau}). The difference
between ZB17b and ZB17c is the treatment of the 
Earth-Moon system (all else being equal, see 
Section~\ref{SecMeth}). The \ems\ option uses
a symplectic sub-integration of the Earth-Moon-Sun 
system, while the \lun\ option is based on a simple 
quadrupole perturbation. One potential explanation
for ZB17b's early separation is that the \ems\ 
option as implemented gives a less accurate 
lunar orbit at constant step size $\D t = 0.375$~d than 
the BS option with adaptive step size control 
(\ems\ and BS both treat the Moon as a separate object).
The \lun\ option (ZB17c) with a correction factor 
$f_{\tt lun} = 0.8525$ \citep{quinn91,varadi03,rauch02}
happens to agree with the BS option (ZB17a) between 
\sm54~Myr and \sm63~Myr. In the following, the \lun\ 
option will be preferred over the \ems\ option.
Note, however, that the relative agreement 
between ZB17a and ZB17c does not prove superior
absolute accuracy of these solutions over others.

\subsection{Orbital solutions ZB17d-p}

The solutions ZB17d-p provide tests of various parameters
(Table~\ref{TabRuns}). The basic setup for all these runs
follows ZB17d using the 2nd-order, 2-day, symplectic 
integrator with \lun\ option. ZB17d diverges from ZB17a
(BS) and ZB17c (the more expensive symplectic version) at 
$\tau \simeq 63$~Myr and $\tau \simeq 68$~Myr, 
respectively (Table~\ref{TabTau}), lending confidence
to the general performance of the ZB17d setup
for $t \ \lsim 63$~Myr. Importantly, the simulations
showed that ZB17e and ZB17d, 
which use initial conditions from INPOP13c and DE431,
respectively, diverge already at \sm54~Myr. It is not
clear at this time which of these ephemerides is 
more accurate. Thus, the
uncertainty in ephemerides currently appears to be one
major limitation for identifying  a unique orbital 
solution beyond \sm54~Myr (see discussion,
Section~\ref{SecDisc}).

ZB17f uses a minimally different orientation for the 
solar rotation axis, which has a minor effect, as 
$\tau \simeq 63$~Myr relative to ZB17d (Table~\ref{TabTau}). 
On the contrary, including only the big 3~asteroids 
(ZB17g) instead of 10~asteroids (ZB17d), drops 
$\tau$ to \sm48~Myr,
indicating a significant influence of asteroids
on the system's dynamic, despite their small mass 
\citep[cf.,][]{laskar11ast}. 
Relative to ZB17d, $\tau$ increases from 48 to 
56~Myr when the number of asteroids ($N$) grows from 3 
to 8, but drops to 54~Myr for $N = 13$ and 16, 
respectively. This represents another major limitation
to finding a unique orbital solution beyond \sm54 Myr
(Section~\ref{SecDisc}).
Tidal dissipation 
in the Earth-Moon system, as well as a hypothetical
Planet~9 appear to have minor effects on the
results (ZB17k and ZB17p, $\tau \simeq 63$ and 
65~Myr, respectively).

Finally, note that using $\tau$ as a criterion, the 
current state-of-the-art solutions (which include 
ZB17a-p but g) all differ from previously published 
results beyond \sm50~Myr (Table~\ref{TabTau}).
The solutions ZB17a-p, as well as La11, diverge from 
La04 at \sm41~Myr. 
At the core of the divergence of the different
orbital solutions lies the Solar System's chaotic 
behavior, i.e., the sensitivity 
to initial conditions and tiny perturbations
(see Section~\ref{SecDisc}). To provide 
some insight into the origin of the
chaos, an eigenmode analysis will be presented 
in the next section. We return to the discussion
of the various orbital solutions in 
Section~\ref{SecDisc}.

\section{Eigenmode analysis} \label{SecEig}

If the mutual planet-planet perturbations were 
sufficiently small (all eccentricities and
inclinations small), then
the full dynamics of the Solar
System could be described by linear secular 
perturbation theory, aka Laplace-Lagrange solution 
\citep[e.g.][]{morbidelli02,malhotra12}.
The existence of chaotic trajectories, however, shows 
that this is not the case. To understand the nature
of higher-order perturbations, it is instructive to examine 
the difference between solutions of the full system 
(numerical) and the linear Laplace-Lagrange 
solution (analytical)
\citep[e.g.][]{applegate86,nobili89,laskar90}.
From the numerical solution of the planets
and Pluto ($i = 1,\ldots,9$), the fundamental 
frequencies were obtained by time-series analysis
of the classical variables:
\beqn
h_i   =   e_i \sin(\vpi_i)        \quad & ; &  \quad
k_i   =   e_i \cos(\vpi_i)        \label{Eqhk} \\
p_i   =   \sin(I_i/2) \sin(\Om_i) \quad & ; &  \quad
q_i   =   \sin(I_i/2) \cos(\Om_i) \label{Eqpq} \ ,
\eeqn
where $e$, $I$, $\vpi$, and $\Om$ are eccentricity, 
inclination, longitude of perihelion, and longitude
of ascending node, respectively. The frequencies
were computed using a zero-padded FFT over the time 
interval 0 to 20 Myr BP (Table~\ref{TabFreqs}) and agree 
well with La10 \citep{laskar11}. The 
largest differences were found for $g_1$ and $s_1$ 
(\sm0\as.008~yr\pmo\ and \sm0\as.005~yr\pmo, 
respectively). The frequency $s_5$ is zero 
because of angular momentum conservation (invariable 
plane).

%---------------------------------------------------------%
%-------------------- TABLE ------------------------------%
%---------------------------------------------------------%
\renewcommand{\baselinestretch}{\blsC}\selectfont
\begin{table*}[ht]
\caption{Fundamental frequencies (arcsec y\pmo) 
and periods (yr) of the Solar System over 20~Myr from 
ZB17c. \label{TabFreqs}}
\vspace*{5mm}
\hspace*{-5mm}
\begin{tabular}{crrrrrrr}
\tableline\tableline
\#  & $g$    & $T_g$      & $s$         & $T_s$    & & $g_{La10}$ $^a$  & $s_{La10}$ $^a$ \\
    & (\asy) & (yr)       & (\asy)      & (yr)     & \hspace*{1cm} & (\asy) & (\asy)      \\
\hline
1 &      5.5821 &    232,170 &   $-$5.6146 &    230,829 & &          5.59 &       $-$5.61\\
2 &      7.4559 &    173,821 &   $-$7.0629 &    183,493 & &         7.453 &       $-$7.06\\
3 &     17.3695 &     74,613 &  $-$18.8476 &     68,762 & &        17.368 &     $-$18.848\\
4 &     17.9184 &     72,328 &  $-$17.7492 &     73,017 & &        17.916 &     $-$17.751\\
5 &      4.2575 &    304,404 &      0.0000 &        $-$ & &      4.257482 &             0\\
6 &     28.2452 &     45,884 &  $-$26.3478 &     49,188 & &       28.2449 &  $-$26.347841\\
7 &      3.0878 &    419,719 &   $-$2.9926 &    433,072 & &      3.087946 &  $-$2.9925258\\
8 &      0.6736 &  1,923,993 &   $-$0.6921 &  1,872,457 & &      0.673019 &    $-$0.69174\\
9 &   $-$0.3494 &  3,709,721 &   $-$0.3511 &  3,691,356 & &    $-$0.35007 &       $-$0.35\\
\tableline
\end{tabular}
\noindent {\small \\[2ex]
$^a$ La10's $g$ and $s$ for comparison \citep{laskar11}.
} 
\end{table*}
\renewcommand{\baselinestretch}{\bls}\selectfont
%---------------------------------------------------------%

It is important to recall that there is no simple one-to-one 
relation
between planet and eigenmode, particularly for the inner 
planets. The system's motion is a superposition of
all eigenmodes, although some modes represent the 
single dominant term for some (mostly outer) planets. 
Assume that each $h,k$ and $p,q$ from
the numerical solution can be approximated
as a linear combination of the eigenmodes 
associated with $g_j$ and $s_j$ plus higher-order 
terms (ellipses):
\beqn
h & \simeq & \sum A_j \sin(g_j t + \Phi_j) + \ldots               \\
k & \simeq & \sum A_j \cos(g_j t + \Phi_j) + \ldots \label{EqhkL} \\
p & \simeq & \sum B_j \sin(s_j t + \Psi_j) + \ldots               \\
q & \simeq & \sum B_j \cos(s_j t + \Psi_j) + \ldots \label{EqpqL} \ ,
\eeqn
where $A_j$ and $B_j$ are amplitudes, and $\Phi_j$ and 
$\Psi_j$ are phases. In the full nonlinear system,
the $g$'s and $s$'s may change over time.
In contrast, in the linear Laplace-Lagrange (LL) solution, 
the fundamental frequencies are constant and higher-order 
terms are absent. Hence a comparison over, say, 100~myr 
of the full solution (Eqs.~(\ref{Eqhk}) and~(\ref{Eqpq}))
vs.\ the linear solution (LL-version of Eqs.~(\ref{EqhkL}) 
and~(\ref{EqpqL})) might provide some insight into the 
chaotic behavior of the full system \citep[e.g.,][]{laskar11}.
LL-version here means no higher terms and constant 
frequencies, amplitudes, and phases (say, obtained 
from a fit over 20~Myr).

For example, in the linear case (denoted by '$^*$' 
in the following; i.e., no higher terms in Eqs.~(\ref{EqhkL}) 
and~(\ref{EqpqL})), we can lump all $h^*$ into a 
vector $\v{h}^*$ ($i = 1,\ldots,9$) and write:
\beqn
\v{h}^* = \v{A} \ \v{u}^* \ ,
\eeqn
where \v{A} is a matrix of amplitudes and $u^*_j = 
\sin(g_j t + \Phi_j)$. For $\det(\v{A}) \neq 0$,
this can be inverted to
give $\v{u}^* = \v{A}\pmo \ \v{h}^*$. For the 
full system, we may write a similar expression
at each time step, $\v{u}(t) = \v{A}\pmo \ \v{h}(t)$. 
However, due to higher-order terms, the amplitudes 
of the $u_j$ will differ from 1 and the frequencies
will no longer be constant. Thus, the deviation
of $\v{u}$ from $\v{u}^*$ (simple sinusoids) provides 
a measure of the importance of the higher-order 
perturbations in the $g$-modes (correspondingly
$\v{v}$ from $\v{v}^*$ with matrix \v{B} in the 
$s$-modes, see $p$-variable above).

%-----------------------------------------------------------%
%--------------------  FIGURE ------------------------------%
%-----------------------------------------------------------%
\renewcommand{\baselinestretch}{\blsC}\selectfont
\begin{figure*}[t]
\def\sc{0.6}
\begin{center}
\includegraphics[scale=\sc]{\figdir 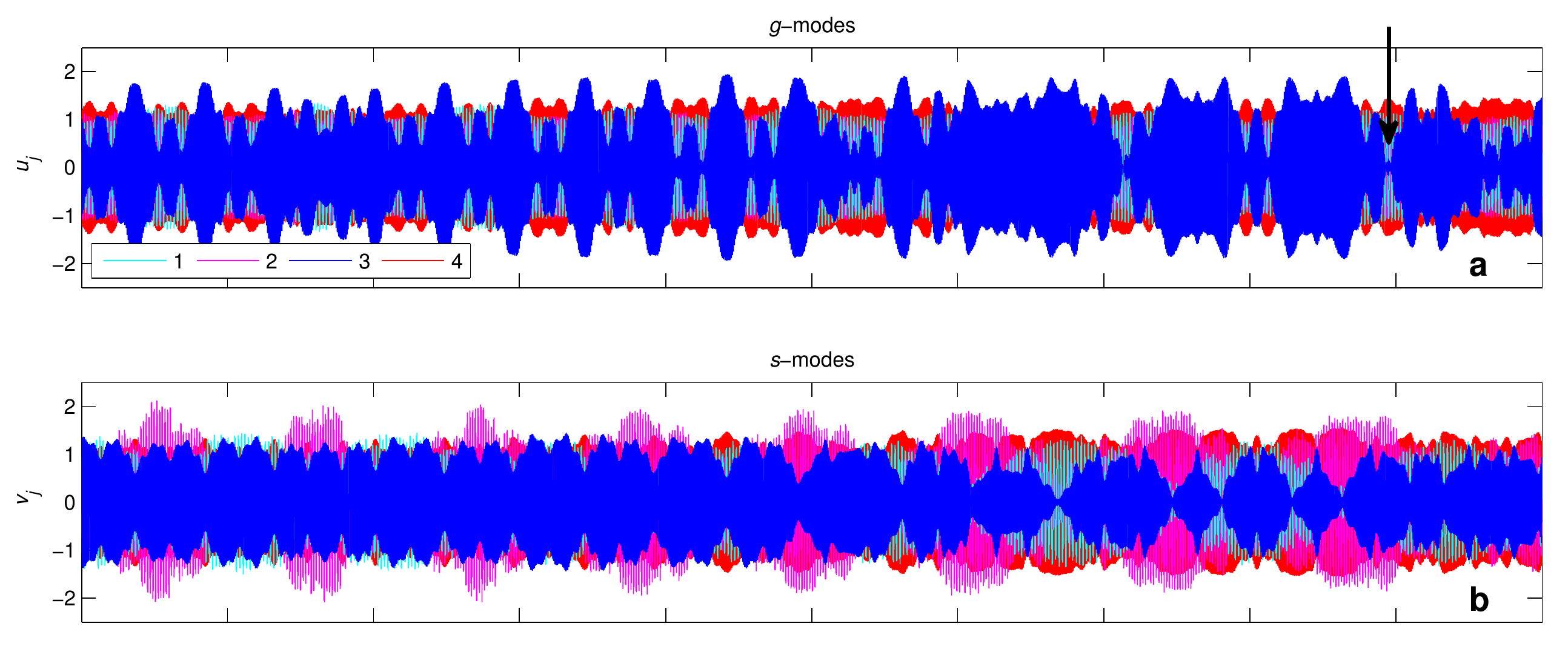}
\hspace*{1mm}
\includegraphics[scale=\sc]{\figdir 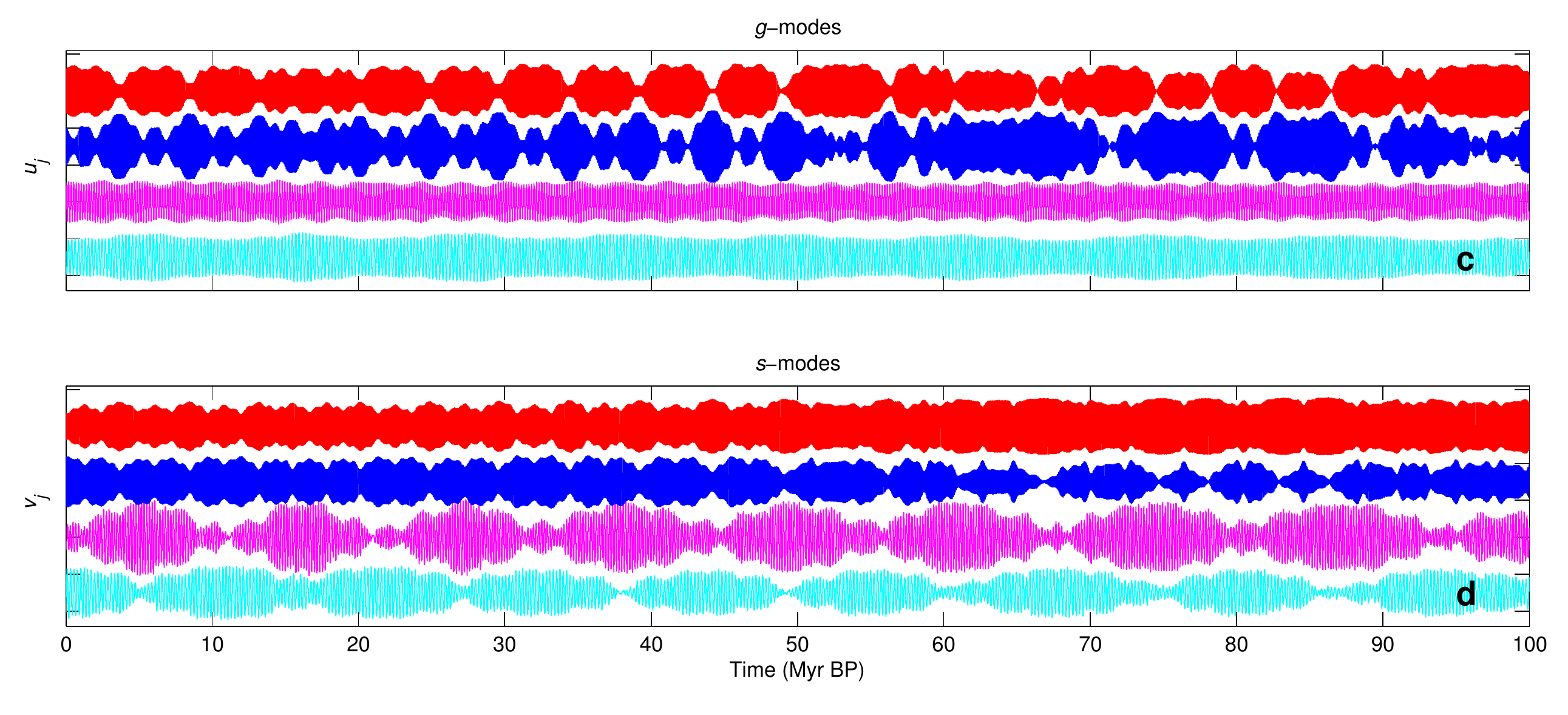}
\end{center}
\caption{\small (a) Eigenmode components
$u_j$ and (b) $v_j$ for $j = 1,\ldots,4$ 
corresponding to $g$- and $s$-modes (for frequencies,
see Table~\ref{TabFreqs}). (c) $u_j$ and 
(d) $v_j$ plotted with arbitrary offsets for clarity.
Note the large 
amplitude variations in e.g., $u_3$, $v_3$ and the change 
in pattern around 50~Myr BP, (cf.\
time of change in secular trends of the arguments 
$\D \Theta_3$ and $\D \Theta_4$, Fig.~\ref{FigArgs}).
The arrow at 89~Myr in (a)
indicates a node in $u_3$ associated with a rapid change
in argument (see Fig.~\ref{FigArgs}).
}
\label{FigUV}
\end{figure*}
\renewcommand{\baselinestretch}{\bls}\selectfont
%-----------------------------------------------------------%

\begin{figure*}[t]
\def\sc{0.55}
\begin{center}
\includegraphics[scale=\sc]{\figdir 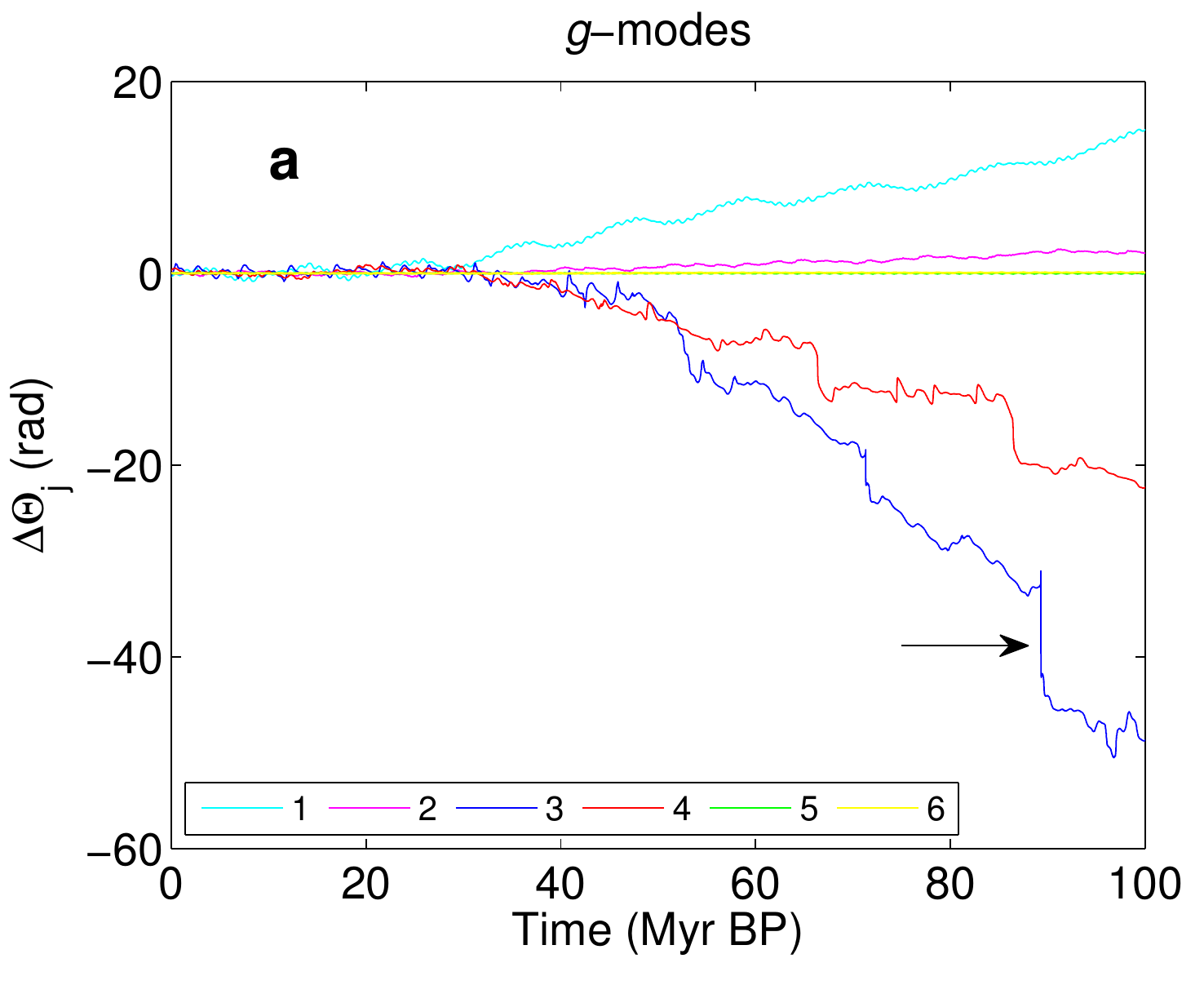}
\includegraphics[scale=\sc]{\figdir 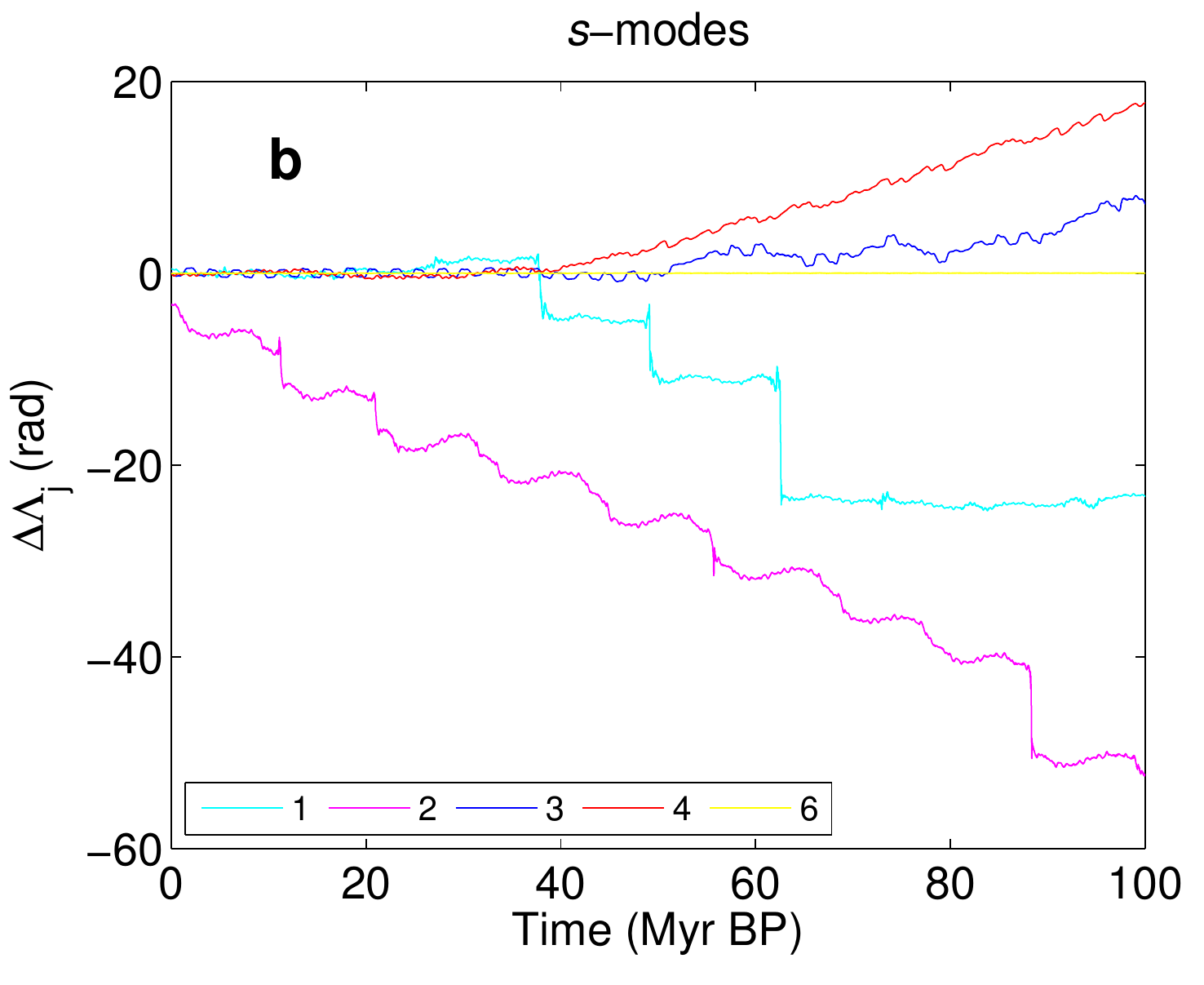}
\end{center}
\caption{\small
Differences in arguments 
(a) $\D \Theta_j  = \Theta_j  - \Theta^*_j$ and
(b) $\D \Lambda_j = \Lambda_j - \Lambda^*_j$
(in radians)
associated with $g$- and $s$-eigenmodes
between the full numerical solution and the linear 
case for $j = 1,\ldots,6$ (except for $s_5 = 0$, see 
text). The arrow at 89~Myr in (a) indicates a rapid 
change in argument associated with a node in $u_3$ 
(see Fig.~\ref{FigUV}). 
}
\label{FigArgs}
\end{figure*}

For $j = 5,\ldots,9$ (dominant in outer planets), 
the \v{u} and \v{v} amplitudes are close to 1 (not 
shown), but not for $j = 1,\ldots,4$ (dominant in 
inner planets, Fig.~\ref{FigUV}). The largest 
amplitude variation in $g$-modes occurs in 
$u_3$ and $u_4$. Also, $u_3$'s and $v_3$'s long-term 
pattern differ between the interval 0-50~Myr vs.\ 50-100~Myr; 
a similar pattern shift occurs in $u_4$ and $v_4$ 
(though not visible in the figure). Such a shift 
is not apparent in $u_1$, $u_2$, $v_1$, and $v_2$.
The largest amplitude variation in $s$-modes 
occurs in $v_2$. As expected, these observations 
suggest that higher-order terms are critical 
for the inner planets. In addition, an apparent
change in eigenmodes occurs around 50~Myr BP in 
the solution ZB17c (as well as in other solutions,
not shown). This point in time corresponds to 
the time of change in secular trends of the arguments 
$\D \Theta_3$ and $\D \Theta_4$ (see Fig.~\ref{FigArgs}).

%-----------------------------------------------------------%
%--------------------  FIGURE ------------------------------%
%-----------------------------------------------------------%
\renewcommand{\baselinestretch}{\blsC}\selectfont
\renewcommand{\baselinestretch}{\bls}\selectfont
%-----------------------------------------------------------%

Changes in the frequencies and phases of the full solution
(constant in the linear case) may be examined by comparing
the arguments of $\v{u}$ and $\v{u}^*$. For the latter, 
we may simply take $g_j t + \Phi_j =: \Theta^*_j$
and $s_j t + \Psi_j =: \Lambda^*_j$
as arguments. For the full solution, a complex variable
will come in handy, which can be defined in the 
linear case as ($i = \sqrt{-1}$):
\beqn
z^*_j = \mu^*_j + i \ u^*_j 
      = \cos(\Theta^*_j) + i \ \sin(\Theta^*_j)
      = e^{i \Theta^*_j} \ ,
\eeqn
where 
$\v{\mu}^* = \v{A}\pmo \ \v{k}^*$ and
$\v{u}^* = \v{A}\pmo \ \v{h}^*$.
By analogy, we compute \v{u}'s arguments from:
\beqn
z_j = \mu_j + i \ u_j 
\eeqn
where
$\v{\mu}(t) = \v{A}\pmo \ \v{k}(t)$ and
$\v{u}(t) = \v{A}\pmo \ \v{h}(t)$. Hence, the 
arguments of the $g$-eigenmodes for the full solution
can be calculated as $\Theta_j = \arctan2(u_j,\mu_j)$. 
In the linear case, the arguments $\Theta^*_j$ 
and $\Lambda^*_j$ simply represent straight lines 
as a function of time
with slopes $g_j$ and $s_j$, respectively. 
Frequency and phase changes in the full solution
will therefore cause deviations from zero in the 
variables
$\D \Theta_j  = \Theta_j  - \Theta^*_j$ and
$\D \Lambda_j = \Lambda_j - \Lambda^*_j$.

%-----------------------------------------------------------%
%--------------------  FIGURE ------------------------------%
%-----------------------------------------------------------%
\renewcommand{\baselinestretch}{\blsC}\selectfont
\begin{figure*}[t]
\def\sc{0.55}
\begin{center}
\includegraphics[scale=\sc]{\figdir 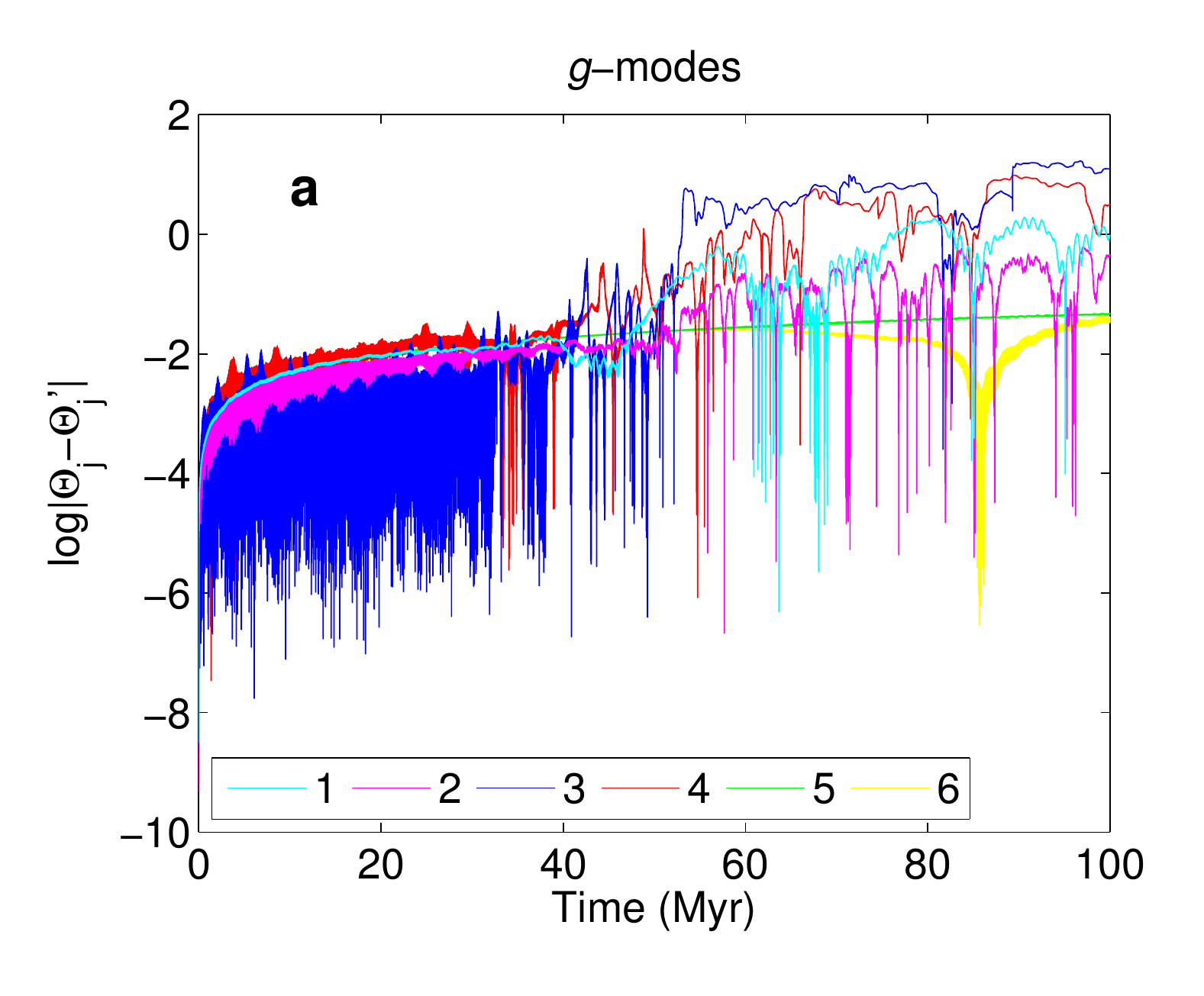}
\includegraphics[scale=\sc]{\figdir 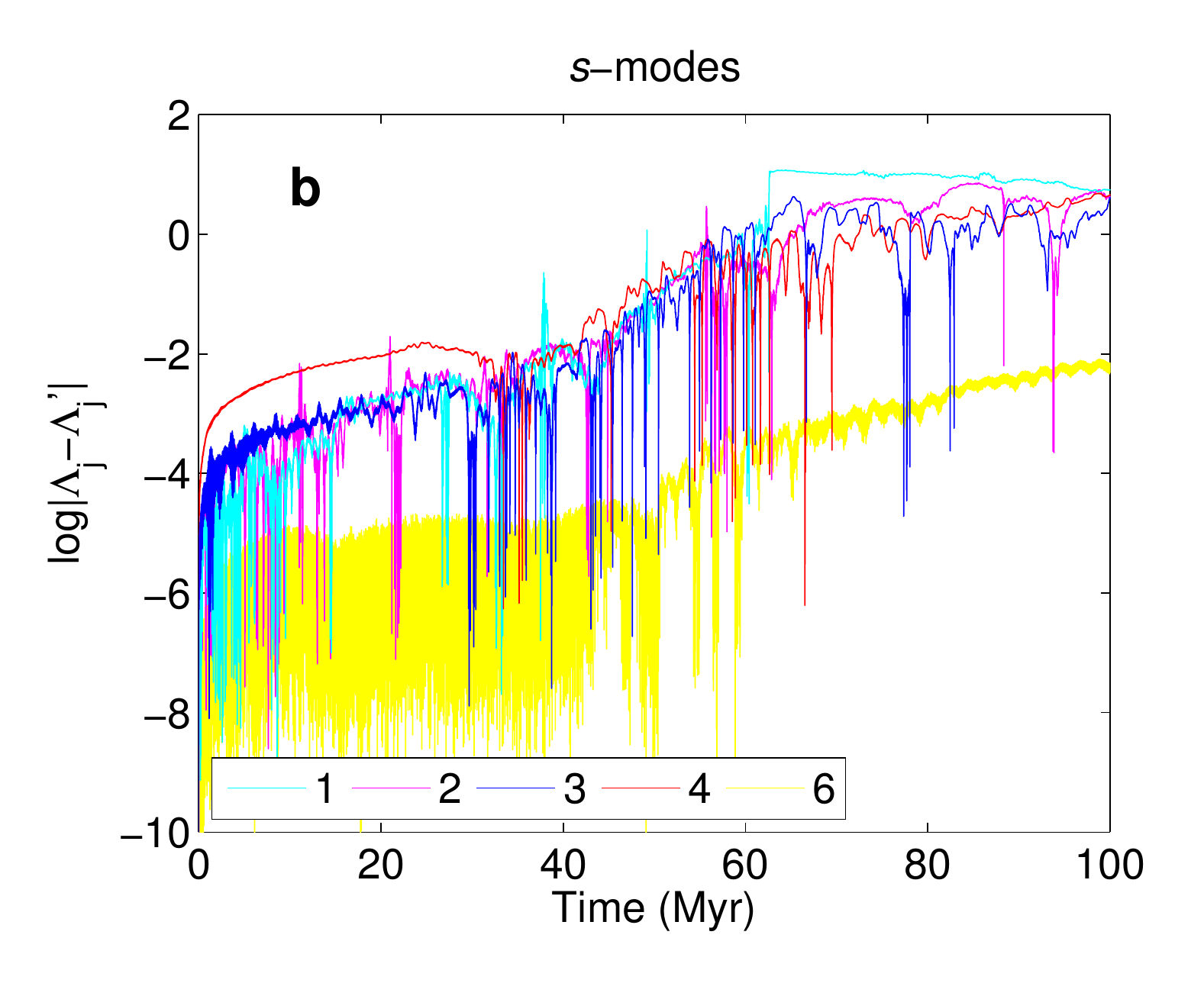}
\end{center}
\caption{\small Example of the difference in arguments 
between two solutions (ZB17b and ZB17c).
(a) $\log | \Theta_j - \Theta'_j |$ and 
(b) $\log | \Lambda_j - \Lambda'_j |$
associated with $g$- and $s$-arguments, respectively.
}
\label{FigDelGS}
\end{figure*}
\renewcommand{\baselinestretch}{\bls}\selectfont
%-----------------------------------------------------------%

The $g$-arguments $\Theta_5$ and $\Theta_6$ 
(dominant in Jupiter and Saturn), show negligible
secular trends over 100~Myr, that is, $g_5$ and $g_6$
are nearly constant, as in the linear case ($\D \Theta_5$ 
and $\D \Theta_6\ \lsim 0.05$~rad, Fig.~\ref{FigArgs}a). 
On the contrary, secular trends in $\D \Theta_3$ and 
$\D \Theta_4$ before \sm40~Myr are typically much larger (for 
various solutions including ZB17c, see Fig.~\ref{FigArgs}a).
For example, $\rd (\D \Theta_3) / \rd t \simeq 30$~rad/50~Myr 
in ZB17c between 40 and 90~Myr (Fig.~\ref{FigArgs}a), 
or $6\e{-4}$~rad~kyr\pmo. If arguments are given as 
$2\pi \hat{g} t$, where $\hat{g}$ is in kyr\pmo\
and $t$ in kyr, then the corresponding frequency 
change is $\D \hat{g} = \rd (\D \Theta)/ \rd t / 2\pi$,
hence $\D \hat{g_3} \simeq 9.5\e{-5}$~kyr\pmo. A FFT
analysis of $\hat{g_3}$ over consecutive 20-Myr intervals 
spanning the full 100~Myr indicates a maximum change of 
\sm8\e{-5}~kyr~\pmo\ in $\hat{g_3}$, corroborating the 
secular trend observed in $\D \Theta_3$. 

The rapid shift
in $\D \Theta_3$ around 89~Myr (Fig.~\ref{FigArgs}a,
arrow), however, is not related to a frequency change.
Such shifts in arguments can occur at the nodes of 
the eigenmodes, where the $u$- or $v$-amplitude
becomes small and the calculated argument changes
rapidly. For example, the $\D \Theta_3$-shift coincides 
with a node in ${u}_3$ at 89~Myr (Fig.~\ref{FigUV},
arrow). In this case, small variations in ${u}_3$'s
and ${\mu}_3$'s amplitudes lead to an apparent rapid 
phase shift between ${u}_3$ and ${\mu}_3$. The offset
in $\D \Theta_3$ between \sm1~Myr prior to, and 
immediately after the shift amounts to $\sm4\pi$; 
otherwise $\D \Theta_3$ remains fairly constant across
the \sm6~Myr interval centered on the shift.
Similar shifts occur in the arguments related to
$g_4$ (Fig.~\ref{FigArgs}a), $s_1$, and $s_2$ 
(Fig.~\ref{FigArgs}b). Before \sm40~Myr, secular 
trends are visible in the arguments $\D \Lambda_3$ 
and $\D \Lambda_4$, associated with $s_3$ and $s_4$.

\subsection{An expression of chaos}

%-----------------------------------------------------------%
%--------------------  FIGURE ------------------------------%
%-----------------------------------------------------------%
\renewcommand{\baselinestretch}{\blsC}\selectfont
\begin{figure}[t]
\def\sc{0.60}
\begin{center}
\includegraphics[scale=\sc]{\figdir 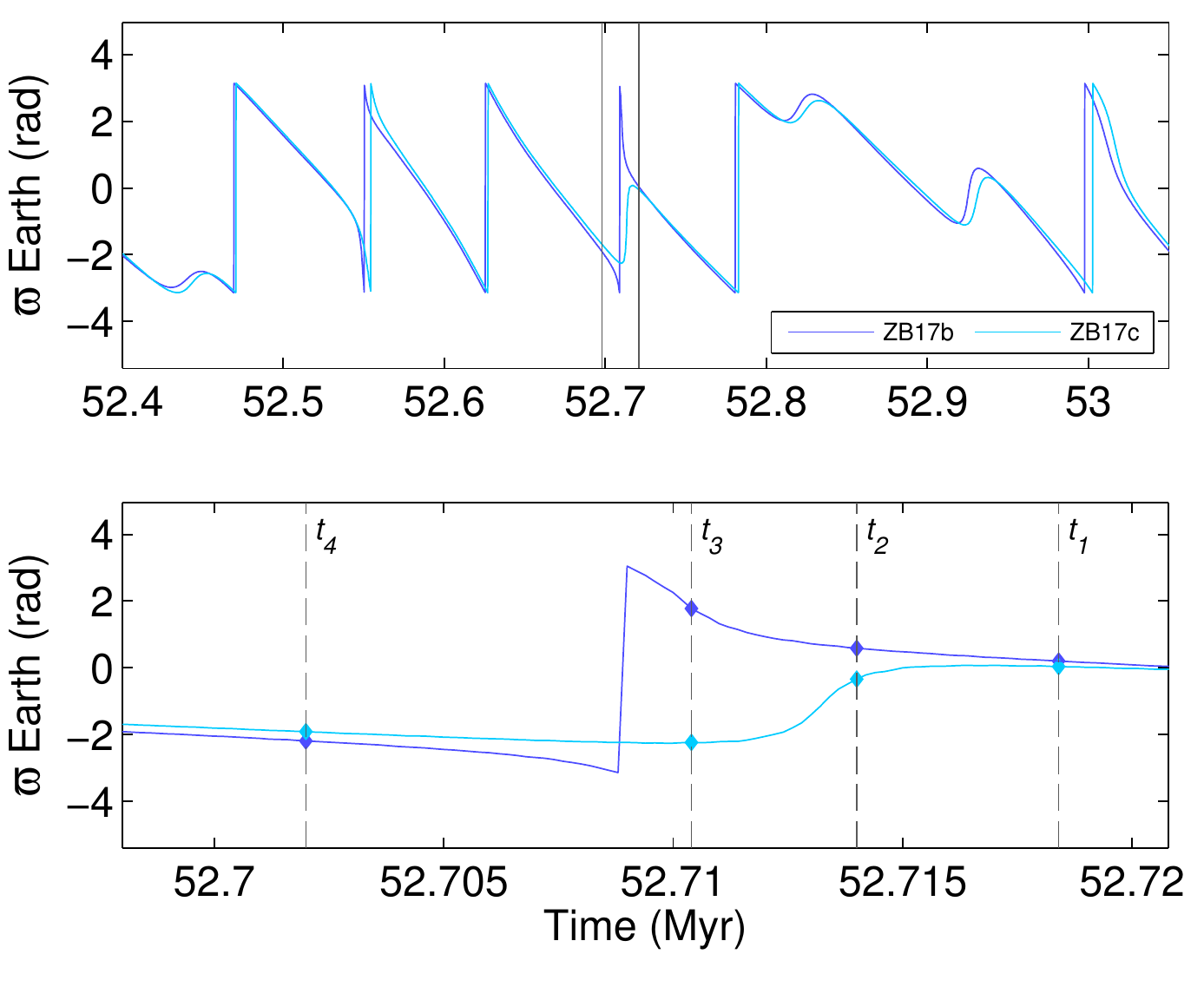}
\end{center}
\caption{\small
Longitude of perihelion $\vpi$ (ZB17b) and $\vpi'$ 
(ZB17c) for Earth's orbit around 52.7~Myr. 
Note different time axes in top and bottom panel.
For illustration of orbits at times $t_1$ to 
$t_4$, see Fig.~\ref{FigOrbs}.
Note that the branching of $\vpi$ and $\vpi'$ for
the two solutions is easy to spot here because 
$\vpi$ crosses from $+\pi$ to $-\pi$ close to $t_3$. 
Otherwise, branching may be more difficult to 
identify but only occurred at times older than $\tau$.
}
\label{FigLongP}
\end{figure}
\renewcommand{\baselinestretch}{\bls}\selectfont
%-----------------------------------------------------------%

Given that the values of $g_3$ and $g_4$ are close
to one another and that both the 
amplitudes and arguments of $u_3$ and $u_4$ show 
the largest variations (and hence deviations from 
the linear solution) suggests that these modes 
are strongly involved in the system's chaotic behavior 
\citep[cf.,][]{laskar90}. It also turned out that 
the difference in $g$-arguments between two different
solutions ($\Theta_j - \Theta'_j$)
grows most rapidly around the eccentricity divergence 
time, $\tau$, for $j = 3,4$ than for other values of $j$.
This is not necessarily the case for $s$-arguments
$\Lambda_j - \Lambda'_j$ (note that divergence 
times for eccentricity and inclination are very 
similar). For ZB17b vs.\ ZB17c, for example, the 
eccentricity-$\tau$ is \sm54~myr. At \sm53~Myr, 
the difference in the $g$-arguments
$\Theta_3 - \Theta'_3$ increases rapidly for 
these two solutions (Fig.~\ref{FigDelGS}a). 

%-----------------------------------------------------------%
%--------------------  FIGURE ------------------------------%
%-----------------------------------------------------------%
\renewcommand{\baselinestretch}{\blsC}\selectfont
\begin{figure}[t]
\def\sc{0.70}
\begin{center}
\includegraphics[scale=\sc]{\figdir 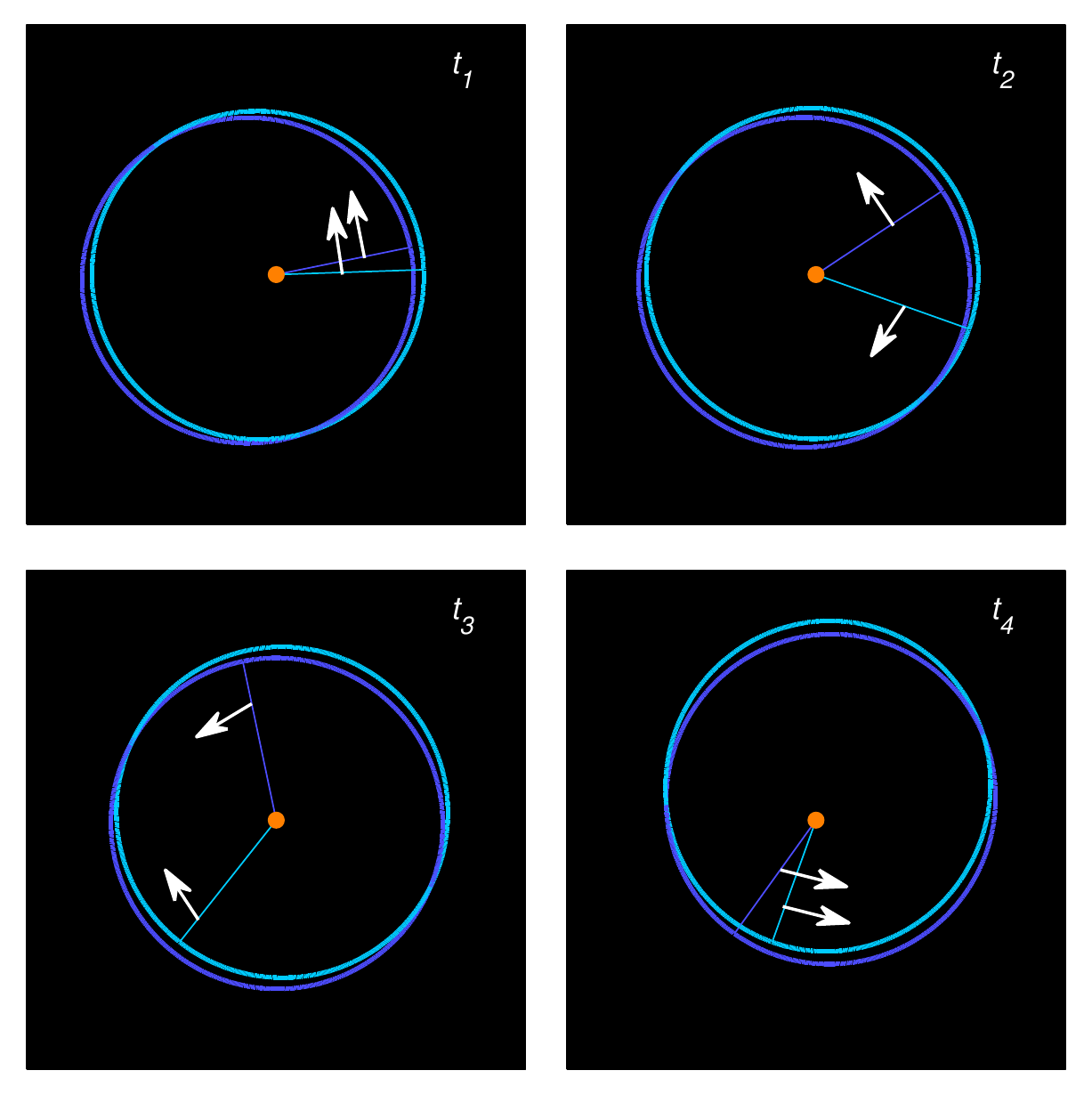}
\end{center}
\caption{\small 
Illustration of Earth's orbit at times $t_1$ to $t_4$
for ZB17b (blue) and ZB17c (cyan)
corresponding to Fig.~\ref{FigLongP}. For visualization,
the eccentricity of Earth's orbit has been 
exaggerated. Straight colored lines fall along lines
of apsides (connecting to perihelia).
Arrows indicate the direction of apse line rotation.
}
\label{FigOrbs}
\end{figure}
\renewcommand{\baselinestretch}{\bls}\selectfont
%-----------------------------------------------------------%

The $g$-modes are related to both the planets' orbital
eccentricities and longitudes of perihelia ($\vpi$'s). 
Examination of $\vpi$ and $\vpi'$ for two different 
solutions close to the divergence time revealed
a first occurrence of branching into two fundamentally 
different physical trajectories for either Venus', 
Earth's, or Mars' orbit.
For example, \lE\ (ZB17b) circulates at 52.71~Myr,
whereas \lEp\ (ZB17c) librates (Fig.~\ref{FigLongP}).
While libration and circulation occur constantly 
across the entire time span, the juncture at 52.71~Myr 
is the first (youngest) occurrence when \lE\ and
\lEp\ take opposite paths ($\vpi$-branching
of the two solutions, illustrated in 
Fig.~\ref{FigOrbs}). Branching occurs frequently 
in the time interval before \sm53~Myr (older). 
In the current example, the first $\vpi$-branching
is described for Earth's orbit and coincides closely
with the rapid rise in $\Theta_3 - \Theta'_3$
at \sm53~Myr (Fig.~\ref{FigDelGS}a). However,
this is not always the case. The first branching
may occur somewhere around $\tau$ for either Venus', 
Earth's, or Mars' orbit. It is likely that the 
preconditioning of $\vpi$ for circulation
vs.\ libration at junctures such as the one
illustrated in Figs.~\ref{FigLongP} and~\ref{FigOrbs}
is sensitive to small differences 
in initial conditions (and/or minuscule perturbations) 
and would therefore represent an expression of the 
chaotic nature of the inner Solar System.

\section{Discussion} \label{SecDisc}

The comparison of the current test solutions 
against published orbital solutions 
(Section~\ref{SecTest}) provides 
some insight into how different numerical realizations 
compare between different investigator groups using 
different codes and integrator packages (external
comparison). The agreement between \citet{varadi03}'s 
R7-run and the current test solution s405 shows 
reproducibility to \sm54~Myr~BP (Fig.~\ref{FigV03}), 
despite the fact that two different
integrator algorithms were used. The comparison 
between \citet{laskar11ast} and the sL11 test-solution 
is slightly less encouraging (Fig.~\ref{FigV03}).
However, when 10~asteroids are included instead of~5, 
the agreement improves (Fig.~\ref{FigZB17}). The 
current state-of-the-art solutions agree with La10$x$ 
($x = $\ a,b,c,d) and La11 up to \sm50~Myr 
(Table~\ref{TabTau}). On the one hand, this is 
encouraging because it suggests validity of the 
solutions over that time period. On the other hand, 
the disagreement beyond \sm50~Myr poses a potential
challenge because the source for the discrepancy is 
unclear at this point. It could reflect a minor issue 
such as small differences in setup parameters
and initial conditions, but could 
also reflect differences in numerical integrators.

The symplectic integrations at different time steps 
(Fig.~\ref{FigTauDt}) showed astonishing consistency,
even at absurdly large step sizes, which leads to at 
least two important conclusions. First, the relationship 
between time step and divergence time is not a robust 
indicator for the absolute accuracy of symplectic 
integrations. Second, the symplectic integration 
with \lun\ option and
12-day time step (5-hours wall-clock time) and the 
4-month Bulirsch-Stoer integration diverge only at 
\sm63~Myr. Thus, full Solar System integrations,
say, for parameter studies over $\lsim$60~Myr may be 
completed within a few hours, rather than months.

The current study provides new state-of-the-art orbital
solutions for applications in geological studies.
\footnote{Numerical solutions are freely available 
at: \myurl}
The solutions ZB17a and ZB17c agree to \sm63~Myr,
despite the fact that two fundamentally different integrator
algorithms were used. The agreement between ZB17a and ZB17c
extends \sm9~Myr beyond that with ZB17b, possibly due to
the \ems\ option used in ZB17b (see Section~\ref{SecABC}).
Also, ZB17a,b,c represent the most 
expensive \BS- and symplectic integrations (smallest error 
per step and smallest symplectic time step, 
Table~\ref{TabRuns}). It is hence conceivable 
that ZB17a and ZB17c are the most accurate solutions 
provided here that are based on DE431 initial conditions
(see discussion below though). However, the relative 
agreement between ZB17a and ZB17c does not prove superior
absolute accuracy of these solutions over others. 
Also, inferring accuracy from step 
size and energy properties of symplectic integrators is
problematic (see Section~\ref{SecNum}).

%-----------------------------------------------------------%
%--------------------  FIGURE ------------------------------%
%-----------------------------------------------------------%
\renewcommand{\baselinestretch}{\blsC}\selectfont
\begin{figure*}[t]
\def\sc{0.48}
\hspace*{-0.5cm}
\includegraphics[scale=\sc]{\figdir 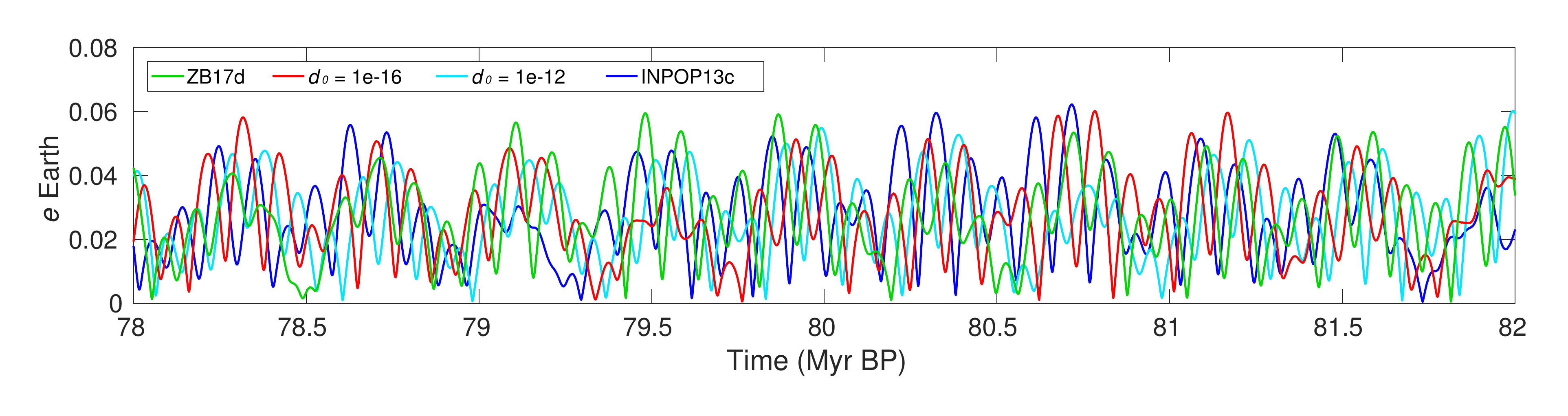}
\caption{\small
Illustration of solutions obtained with the same
setup and integrator as ZB17d (DE431), except for small 
differences in initial conditions. Time interval shown 
is sufficiently beyond divergence time. Values for
\dn\ indicate difference to DE431 in Earth's 
$x$-coordinate. Initial conditions from INPOP13c 
were applied to all planets and Pluto.
}
\label{Figd0}
\end{figure*}
\renewcommand{\baselinestretch}{\bls}\selectfont
%-----------------------------------------------------------%

Given the chaotic nature of the system (see Section~\ref{SecEig}),
the uncertainty in ephemerides (e.g., difference between
DE431 and INPOP13c) currently appears to be one major 
limitation for identifying  a unique orbital solution beyond 
\sm54~Myr.
The number of asteroids ($N$) included in the simulations 
represents another major limitation. One might expect that 
once a certain number of asteroids has been included, the
divergence time $\tau$ would remain constant,
as the effect of asteroid mass on the 
system's dynamic would approach a limit. However, this
is not the case. For example, while relative to ZB17d
($N = 10$), $\tau$ does increase from 48 to 56~Myr when 
$N$ grows from 3 to 8, $\tau$ drops to 54~Myr 
for $N = 13$ and 16, respectively (Table~\ref{TabTau}). 
Given $\tau = 54$~Myr, asteroids therefore represent a similarly 
limiting factor as initial conditions.

\subsection{Constraints from geologic records}

Can geologic evidence 
help to constrain astronomical solutions further back 
in time? For example, one approach currently pursued 
is to search for and identify chaotic resonance transitions 
in geologic records (occurring at specific ages), which
would then have to be matched by a certain orbital 
solution that shows a resonance transition at about the 
same age \citep[e.g.,][]{paelike04,mameyers17}.
Ignoring all other physical and numerical limitations
(see, e.g.,  Table~\ref{TabRuns}), what would a 
more systematic search for matching solutions
entail in practical terms, for now only focusing
on initial conditions as a source of uncertainty?

The difference in Earth's initial position $\v{x}_0 = 
(x_0,y_0,z_0)$ between DE431 and INPOP13c in each
coordinate is \sm$10^{-9}$~AU (= \dn, units
omitted hereafter). Values within the interval, say 
$x_0 + \dn$, may hence be selected as new $x'_0$, 
each of which will lead to a different numerical, 
orbital solution (ensemble of $K$ total solutions). 
Numerically, $K$ may then be estimated as follows 
(other mathematical and physical considerations 
aside). At double precision, as used here, the 
machine epsilon ($\eps$) is of order $10^{-16}$. 
Thus, if $x_0$ is of order 1,
a set of ensemble initial conditions for $x'_0$
may be selected as $x'_0 = x_0 + n \cdot \eps$, where $n =
1,\ldots,K$. Hence $K = \dn/\eps = 10^{7}$, which,
combined with sets of $y'_0$ and $z'_0$, would give $10^{21}$
possible initial values just for Earth's position 
that could be used to generate an ensemble of numerical 
test solutions. In addition, similar estimates can be 
made for velocities and for all other bodies of 
the Solar System, which would give a very large 
number of potential initial conditions --- clearly 
too large for practical analysis.

Just to illustrate a first step of such an approach, 
consider changing only Earth's $x_0$ by $1\e{-16}$,
for instance, in the ZB17d setup (which uses DE431). 
This gives a solution with divergence time $\tau 
\simeq 57$~Myr that bears no resemblance to ZB17d, 
for example, around 80~Myr, except for the 
ubiquitous 405-kyr cycle ($g_2-g_5$), which is 
omnipresent in all solutions (Fig.~\ref{Figd0}).
Adding another solution with $\dn = 1\e{-12}$
and ZB17e (based on INPOP13c), graphically 
illustrates several major difficulties when
attempting to identify solutions with certain
properties in a system with chaotic behavior
(Fig.~\ref{Figd0}). First, and unsurprisingly,
for times sufficiently beyond $\tau$, the solutions
do not show any systematic pattern or behavior
as a function of the size of \dn. That is, the 
properties of the solutions at that point
appear random, regardless of whether 
$\dn = 1\e{-16}$, or $1\e{-12}$, etc.
Second,
divergence times appear similarly arbitrary. For 
example, for $\dn = 1\e{-16}$ and $1\e{-12}$,
$\tau \simeq 57$~Myr and \sm63~Myr, respectively.
In other words, the solution with the larger offset 
in initial conditions shows 'better' (extended) 
agreement with ZB17d in the long run. 

The behavior described above is characteristic for chaotic 
systems and merely highlights the obstacles in tracking
solutions with different properties and different initial 
conditions (cf.\ eigenmode analysis, Section~\ref{SecEig}). 
As a result, even if it is possible to 
identify, say, resonance transitions in geological 
sequences, it is not obvious at this point how this
information can be used to pinpoint a unique numerical 
orbital solution. Given the vast number of possible
initial conditions, it is likely that a large 
number of solutions can be generated that will match 
the geological observations within data uncertainty.
Conversely, can solutions at least be singled 
out and excluded that do not match the observations?
For such an effort to be successful, generation of
long, highly-quality, continuous geologic records 
that unequivocally identify resonance transitions 
should be a high priority.

\section{Conclusions} \label{SecConcl}

The results of the present integrations lead to several
conclusions regarding the factors that currently limit 
the identification
of a unique orbital solution beyond \sm50~Myr. In the following,
the factor with the smallest divergence time $\tau$ is 
considered to be currently limiting, those factors with 
larger $\tau$'s are not (Table~\ref{TabTau}). If we prefer
the \lun\ over the \ems\ option, then the choice of the numerical 
algorithm at the precision tested here is not limiting; 
the \BS\ and symplectic run with \lun\ option 
agree to \sm63~Myr. Note also
that $\tau$ is not dominated by integration errors 
(Fig.~\ref{FigZB4d-2d}), hence using, e.g., extended or
quadruple instead of double precision is unlikely to 
affect divergence times \citep[cf.][]{laskar11}. Moreover,
the step size of symplectic integrations (\lun\ option)
appears much less critical than one may think
and might be increased to up to \sm12~days for some 
applications (see Fig.~\ref{FigTauDt}). Integrations 
over 100~Myr could then be run in a few hours, rather than 
months. The potential perturbation of a hypothetical 
Planet~9 on Earth's orbit as tested here is not a limiting 
factor ($\tau \simeq 65$~Myr).
Currently, the limiting factors ($\tau \simeq 54$~Myr) 
appear to be the number of asteroids included and
uncertainties in initial conditions for 
positions and velocities of Solar System bodies as given 
by NASA and IMCCE. Overcoming these limitations should therefore
be the focus of future research in order to push the limits 
of an astronomically-tuned geologic time scale further
back in time. However, given the fundamental barriers 
discussed in the previous section, the path toward
achieving this goal is not obvious.

\vfill
\noindent
{\small
{\bf Acknowledgments.}
I thank the anonymous reviewer for comments, which improved
the manuscript.
I am grateful to Bruce Runnegar and Michael Ghil for
providing the numerical output of the simulations from 
\citet{varadi03}, also available at: \myurl.
}

\software{\hnb\ \citep{rauch02},
SPICE (\url{naif.jpl.nasa.gov/naif/toolkit.html}),
calceph (\url{www.imcce.fr/inpop/calceph}), Matlab.}

%--------------------  BIB  --------------------------------%
% (1) run bibtex. (2) cp ZeebeAJ17P.bbl ZeebeAJ17P.bib
%
%\bibliography{../../../latex/rz}
\bibliographystyle{corres/apj}
%
% cp ZeebeAJ17P.bib ZeebeAJ17P.bbl

\end{document}